\documentclass[iop]{emulateapj-rtx4}
\slugcomment{{\sc Accepted to ApJ:} 2017 February 14}

\newcommand{\kms}{$\rm km\, s^{-1}$}

 \shortauthors{French $\&$ Wakker}
\usepackage{graphicx}
\usepackage{subfigure}

\begin{document}

\title{Probing Large Galaxy Halos at $z\sim0$ with Automated $\rm L \MakeLowercase{y} \alpha$-Absorption Matching}

\author{David M. French and Bart P. Wakker}

\affil{Department of Astronomy, University of Wisconsin, Madison, WI 53706, USA}

\begin{abstract}

We present initial results from an ongoing large-scale study of the circumgalactic medium in the nearby Universe ($cz \leq 10,000$ $\rm km\, s^{-1}$), using archival Cosmic Origins Spectrograph spectra of background quasi-stellar objects. This initial sample contains 33 sightlines chosen for their proximity to large galaxies ($D\geq25$ kpc) and high signal-to-noise ratios (S/N $\geq$ 10), yielding 48 Ly$\alpha$ absorption lines that we have paired with 33 unique galaxies, with 29 cases where multiple absorbers within a single sightline are paired with the same galaxy. We introduce a likelihood parameter to facilitate the matching of galaxies to absorption lines in a reproducible manner. We find the usual anti-correlation between Ly$\alpha$ equivalent width ($EW$) and impact parameter ($\rho$) when we normalize by galaxy virial radius ($R_{vir}$). Galaxies associated with a Ly$\alpha$ absorber are found to be more highly-inclined than galaxies in the survey volume at a $>99\%$ confidence level (equivalent to $\sim 3.6 \sigma$ for a normal distribution). In contrast with suggestions in other recent papers of a correlation with azimuth angle for Mg\,{\sc ii} absorption, we find no such correlation for Ly$\alpha$.

\end{abstract}

\keywords{galaxies:intergalactic medium, galaxies:evolution, galaxies:halos, quasars: absorption lines}

\section{INTRODUCTION}

It is well known that galaxies must continue to accrete gas throughout their lifetimes in order to sustain their observed levels of star formation (e.g., Erb 2008; Prochaska $\&$ Wolfe 2009; Putman et al. 2009a, 2009b; Bauermeister et al. 2010; Genzel et al. 2010). This additional gas must come from the diffuse intergalactic medium (IGM), where the majority of the baryons in the universe reside (Penton et al. 2002, 2004; Lehner et al. 2007; Danforth $\&$ Shull 2008; Shull et al. 2012). How exactly this IGM gas eventually falls into the halos and disks of galaxies is still highly uncertain, as observational constraints are hard to come by. Because of the diffuse nature of IGM gas, it is most readily and sensitively detected as absorption in the spectra of background quasi-stellar objects (QSOs). The advent of the sensitive ultraviolet (UV) Cosmic Origins Spectrograph (COS) on the \textit{Hubble Space Telescope} (\textit{HST}; Osterman et al. 2011; Green et al. 2012) has provided a wealth of information about the properties and distribution of both the ions of heavy elements as well as the Lyman series of neutral hydrogen (H\,{\sc i}) gas around galaxies. 

Individual concentrations of gas along a given sightline imprint absorption lines onto the spectrum in the direction of the QSO. The metal lines trace the star formation history within the intervening gas, and H\,{\sc i} lines (e.g., Ly$\alpha$) indicate both the location and velocities of outflowing gas, as well as the presence of fuel for future star formation. Numerous studies using these observations have shown that many Ly$\alpha$ absorbers trace individual galaxy halos (e.g., Lanzetta et al 1995; Chen et al. 1998, 2001; Tripp et al. 1998; Wakker $\&$ Savage 2009; Steidel et al. 2010; Prochaska et al. 2011; Tumlinson et al. 2011, 2013; Thom et al. 2012;  Stocke et al. 2013, 2014; Liang $\&$ Chen 2014; Tejos et al. 2014; Borthakur et al. 2015).

Some recent studies found that about half of Ly$\alpha$ absorbers lie within galaxy halos, at impact parameters $\rho<350$ kpc (C\^{o}t\'{e} et al. 2005; Prochaska et al. 2006; Wakker $\&$ Savage 2009). In addition, Wakker $\&$ Savage (2009) found that an absorber lies within 400 kpc and 400 $\rm km\, s^{-1}$ for $90\%$ of galaxies brighter than $0.1L_{\**}$, and all galaxies have a Ly$\alpha$ absorber within 1.5 Mpc. Higher-redshift studies, such as Rudie et al. (2012a) at $2<z<3$, found evidence for an elevated density of absorbers up to 2 Mpc from galaxies. Wakker $\&$ Savage (2009) also discovered a correlation between Ly$\alpha$ absorption linewidth and impact parameter $\rho$, observing that the broadest lines (FWHM $>$150 $\rm km\, s^{-1}$) are only seen within 350 kpc of a galaxy, while at $\rho>1$ Mpc, only lines with FWHM $<75$ $\rm km\, s^{-1}$ occur. This suggests that the temperature and/or turbulence of gas increases in the presence of galaxies, a hypothesis that has been further supported by the results of Wakker et al. (2015). 

Studying the enrichment of galaxy halos is necessary for constraining outflow models and informing stellar feedback prescriptions. Directly measuring the velocities and column densities of absorbers as a function of impact parameter and orientation around galaxies would provide the clearest evidence of inflow or outflow activity, but results are still uncertain. Kacprzak et al. (2011b) claimed to find that Mg\,{\sc ii} equivalent widths correlate with galaxy inclination, but Mathes et al. (2014) found no such correlation for Ly$\alpha$ and O\,{\sc vi} absorbers. Furthermore, we should expect outflowing gas to be more highly enriched and trace the metallicity of the associated galaxy, with inflowing gas instead appearing only in H\,{\sc i}. Both Stocke et al. (2013) and Liang $\&$ Chen (2014) found an ``edge'' to heavy ion absorption at $\sim0.5R_{vir}$, but found Ly$\alpha$ covering fractions of $\sim0.75-1$ continuing out to $R_{vir}$. However, Mathes et al. (2014) measured O\,{\sc vi} absorption out to $\sim3 R_{vir}$, and Savage et al. (2014) found that more than half of O\,{\sc vi} absorption occurs beyond 1 $R_{vir}$ from the nearest galaxy. Additionally, Borthakur et al. (2015) found that Ly$\alpha$ absorption $EW$ correlates with galaxy H\,{\sc i} gas fraction, but only weakly with SFR, suggesting that accretion flow from the CGM is slow and continuous.

Recent results from Kacprzak et al. (2011b, 2012a) suggest that absorbing systems have a preferred orientation with respect to the major and minor axes of the galaxies they are associated with. This could be evidence of inflows and outflows, or an effect of the global structure of galaxy halos, but the statistics are not yet good enough to provide consistent answers. A larger-scale study of inclination and azimuthal angles versus absorber properties is needed in order to elucidate the distribution of absorbing systems around galaxies. This is most easily done for the largest galaxies in the nearby universe, where it is possible to obtain inclinations and unambiguous absorber associations. 

Previous studies have suffered from small sample sizes (e.g., Mathes et al. 2014 used 14 galaxies, Stocke et al. 2013 used 11, Werk et al. 2014 used 44), incompleteness due to their higher mean redshifts (e.g., the Mathes et al. 2014 sample is $0.12 <z<0.67$), and limited impact parameter reach (e.g., Werk et al. 2014 probed CGM gas only within $\rho < 160$ kpc of galaxies). To address these shortcomings, we are conducting a large survey of the properties of intergalactic gas in the nearby universe, where we have good and relatively complete information on both faint and bright galaxies, in order to reveal how the IGM and galaxies affect each other. 

We are taking advantage of the over 500 archived QSO and Seyfert spectra taken by the COS and Space Telescope Imaging Spectrograph (STIS) on \textit{HST}, combined with the wealth of information available for the $\sim100,000$ galaxies with $cz<10,000$ $\rm km\, s^{-1}$ found in the NASA Extragalactic Database (NED) to probe the environment of absorbing gas systems in the nearby universe. In this paper we introduce a new, numerical method for associating absorption lines with nearby galaxies. This approach will allow for an objective understanding of the distribution of the gas around galaxies, which requires looking for both detections and non-detections of gas, both near and far from galaxies, with a robust and reproducible metric for matching galaxies with absorption.

This paper presents our likelihood-matching method with initial results from a pilot study of 33 sight lines, chosen for their proximity to large galaxies and high signal-to-noise spectra. It is organized as follows. In Section 2 we present the data and analysis techniques, in Section 3 we present the results, in Section 4 we discuss possible interpretations of our results, and in Section 5 we present a summary.

\begin{table*}[ht]
\caption{\small{COS Targets in this Sample}}
  \vspace{-15pt}
\begin{center}
\begin{tabular}{l l l l l l l l l l}
 \hline \hline
  Target 					& R.A. 		& Decl. 		 & \textit{z}	  & Program 	  & Grating 	   & Obs ID 	    & Obs Date 	     & $\rm T_{exp}^a$     & $\rm S/N^a$  \\ 
  	    					& 	       		&	  		 & 		  	  & 		    	  & 		  	   & 		  	    & 		     	     & 	        [ks]         & [1238] \\ 
 \scriptsize (1)  				& \scriptsize (2) & \scriptsize (3) & \scriptsize (4) & \scriptsize (5) & \scriptsize (6) & \scriptsize  (7) & \scriptsize (8) & \scriptsize (9) & \scriptsize (10)  \\ \hline \hline
\\
    
1H0717+714		  		&  07 21 53.3  &  +71 20  36	&    0.5003	& 12025	&   G130M	&   LBG812	& 11-12-27      	 	  &  6.0    &      37         \\
2dFGRS\_S393Z082  		&  02 45 00.8  &  $-$30 07  23	&    0.3392	& 12988	&   G130M	&   LC1040	& 13-05-27,28 	 	  & 17.7   &      10         \\
				  		&		       &			&			&		&			&    LC1045	&				  &	       &		   \\
FBQSJ1431+2442     		&  14 31 25.8  &  +24 42 20	&   0.4069		& 13342	&   G130M	&   LC8903	& 15-03-29		  & 16.5  &      17          \\
				 		&		       &			&			& 12603	&			&   LBS314	& 13-03-08		  &	      &		  	  \\
H1101-232   		 		&  11 03 37.7  &  $-$23 29 31	&   0.1860		& 12025	&   G130M	&   LBG804	& 11-07-05  		   & 13.3  &      16         \\
HE0241-3043  		 		&  02 43 37.7  &  $-$30 30 48	&   0.6693		& 12988	&   G130M	&   LC1070	& 13-06-21  		   & 7.0    &      14         \\
LBQS1230-0015  	 		&  12 33 04.1  &  $-$00 31 34    	&   0.4709		& 11598	&   G130M	&   LB5N15	& 10-08-01  		   & 10.3  &      13         \\
				 		&		       &			&			& 12486	&			&   LBP250	& 12-04-26		   &	       &	  	   \\
MRC2251-178  	 		&  22 54 05.9  &  $-$17 34 55	&   0.0661		& 12029	&   G130M	&   LBGB03	& 11-09-29                   &  5.5   &      42         \\
Mrk290  					&  15 35 52.3  &  +57 54 09	&   0.0296		& 11524	&   G130M	&   LB4Q02	& 09-10-28  		   &   3.9  &      38         \\
Mrk876  					&  16 13 57.2  &  +65 43 10	&   0.1290		& 11524	&   G130M	&   LB4Q03	& 10-04-08,09  		   & 12.6  &      65         \\
				 		&		       &			&			& 11686	&			&   LB4F05	&				   &	       &	  	   \\
Mrk1014  					&  01 59 50.2  &  +00 23 41	&   0.1630		& 12569	&   G130M	&   LBP404	& 12-01-25  		   &  1.8   &      17         \\
PG0832+251  				&  08 35 35.9  &  +24 59 41	&   0.3310		& 12025	&   G130M	&   LBG808	& 12-04-19		   &  6.1   &      14         \\
PG0003+158  				&  00 05 59.3  &  +16 09 49	&   0.4509		& 12038	&   G130M	&   LBGL17	& 11-10-22  		   & 10.4  &      25         \\
PG1001+054  				&  10 04 20.1  &  +05 13 01	&   0.1610		& 13347	&   G130M	&   LCCV02	& 14-06-19  		   &  5.2   &      14         \\
						&		       &			&		 	& 13423	&			&   LC9W02	& 14-04-04		   &	       &	  	   \\
PG1302-102  				&  13 05 33.0  &  $-$10 33 20	&   0.2784  	& 12038	&   G130M	&   LBGL04	& 11-08-16  		   &  6.0   &      27         \\
RBS1768  				&  21 38 49.7  &  $-$38 28 40	&   0.1830  	& 12936	&   G130M	&   LC1201	& 13-06-25		   &  7.0   &      24         \\
RX J0714.5+7408  			&  07 14 36.2  &  +74 08 11	&   0.3710  	& 12275	&   G130M	&   LBH402	& 11-03-18   		   &  8.3   &      18         \\
RX J1017.5+4702  			&  10 17 30.9  &  +47 02 25	&   0.3354  	& 13314	&   G130M	&   LC9M04	& 14-01-29		   &  8.7   &      12         \\
RX J1117.6+5301  			&  11 17 40.5  &  +53 01 50	&   0.1587  	& 14240	&   G130M	&   LCWM05	& 16-04-13		   &  4.9   &      11         \\
RX J1236.0+2641  			&  12 36 04.1  &  +26 41 36	&   0.2092  	& 12248	&   G130M	&   LBH087	& 12-01-29  		   &  4.2   &      11         \\
RX J1330.8+3119  			&  13 30 53.2  &  +31 19 32	&   0.2423  	& 12248	&   G130M	&   LBHO85	& 11-07-11		   &   4.3  &      11         \\
RX J1356.4+2515  			&  13 56 25.6  &  +25 15 23	&   0.1640  	& 12248	&   G130M	&   LBH057	& 12-02-03		   &   2.3  &      10         \\
RX J1503.2+6810  			&  15 03 16.5  &  +68 10 06	&   0.1140  	& 12276	&   G130M	&   LBI609		& 10-12-31 		   &   1.9  &      11         \\
RX J1544.5+2827  			&  15 44 30.5  &  +28 27 56	&   0.2314  	& 13423	&   G130M	&   LC9W08	& 14-02-25		   &   2.1  &      10         \\
RX J2043.1+0324  			&  20 43 06.2  &  +03 24 50	&   0.2710  	& 13840	&   G130M	&   LCJW02	& 14-10-23		   &   7.8  &      15         \\
RX J2139.7+0246  			&  21 39 44.2  &  +02 46 05	&   0.2600  	& 13840	&   G130M	&   LCJW03	& 14-10-27  		   & 	7.9  &      16         \\
SBS0957+599         			&  10 01 02.6  &  +59 44 15	&   0.7475  	& 12248	&   G130M	&   LBHO65	& 11-03-18,19  		   &   3.3  &      12         \\
SDSSJ021218.32-073719.8  	&  02 12 18.3  &  $-$07 37 20	&   0.1739  	& 12248	&   G130M	&   LBHO83	& 11-06-26		   &   6.5  &      12         \\
				        	     	& 	  	       &			&    	  	 	&		&			&   LBHO92	& 11-08-21		   &          &                   \\
				        	     	& 	               &			&    	  	 	&		&			&   LBHO92	& 11-08-21		   &          &                   \\
SDSSJ080838.80+051440.0 	&  08 08 38.8  &  +05 14 40	&   0.3606  	& 12603	&   G130M	&   LBS330	& 12-03-17  		   &   4.7  &      10         \\
SDSSJ091728.60+271951.0 	&  09 17 28.6  &  +27 19 51	&   0.0756  	& 14071	&   G130M	&   LCX202	& 15-11-30		   & 15.5  &      11         \\
				        	     	& 	  	      &				&    	  	 	&		&			&   LCX2Z2	& 16-02-06		   &          &                   \\
SDSSJ112224.10+031802.0 	&  11 22 24.1  &  +03 18 02	&   0.4753  	& 12603	&   G130M	&   LBS318	& 13-03-29		   &   7.6  &      13         \\
SDSSJ130524.30+035731.0 	&  13 05 24.3  &  +03 57 31	&   0.5457  	& 12603	&   G130M	&   LBS321	& 12-06-25,26		   &   7.6  &      13         \\
SDSSJ135726.27+043541.4 	&  13 57 26.2  &  +04 35 41	&   1.2345  	& 12264	&   G130M	&   LBJ005	& 11-06-22		   & 14.1  &      21         \\
				        	     	& 	  	      &				&    	  	 	&		&			&   LBJ007	& 11-06-26		   &          &                   \\
SDSSJ140428.30+335342.0 	&  14 04 28.3  &  +33 53 42	&   0.5500  	& 12603	&   G130M	&  LBS320	& 13-03-03   		   &   7.7  &      10          \\
TON1009  			    	&  09 09 06.1  &  +32 36 31	&   0.8103  	& 12603	&   G130M	&  LBS328	& 12-04-22   		   &   4.7  &      12         \\
 \\
\hline
\end{tabular}

\footnotesize \raggedright \textbf{Note.}\\
\footnotesize \raggedright $^{\rm a}$ Total exposure time and S/N ratio is given for multi-orbit exposures. \\
\end{center}
  \label{table1}
\end{table*}

\section{DATA AND ANALYSIS}

\subsection{Galaxy Data}
Achieving the goal of this study relies on knowing the locations and properties of all galaxies near detected Ly$\alpha$ absorption lines. To facilitate this, we have constructed a database of all $z\leq 0.033$ ($cz\leq 10,000$ $\rm km\, s^{-1}$) galaxies with published data available through the NASA Extragalactic Database (NED). A full description of this catalog will be presented in D. M. French $\&$ B. P. Wakker (2017, in preparation). Here we summarize its most important aspects.

\begin{figure}[b!]
        \centering
        \vspace{0pt}
        \includegraphics[width=0.50\textwidth]{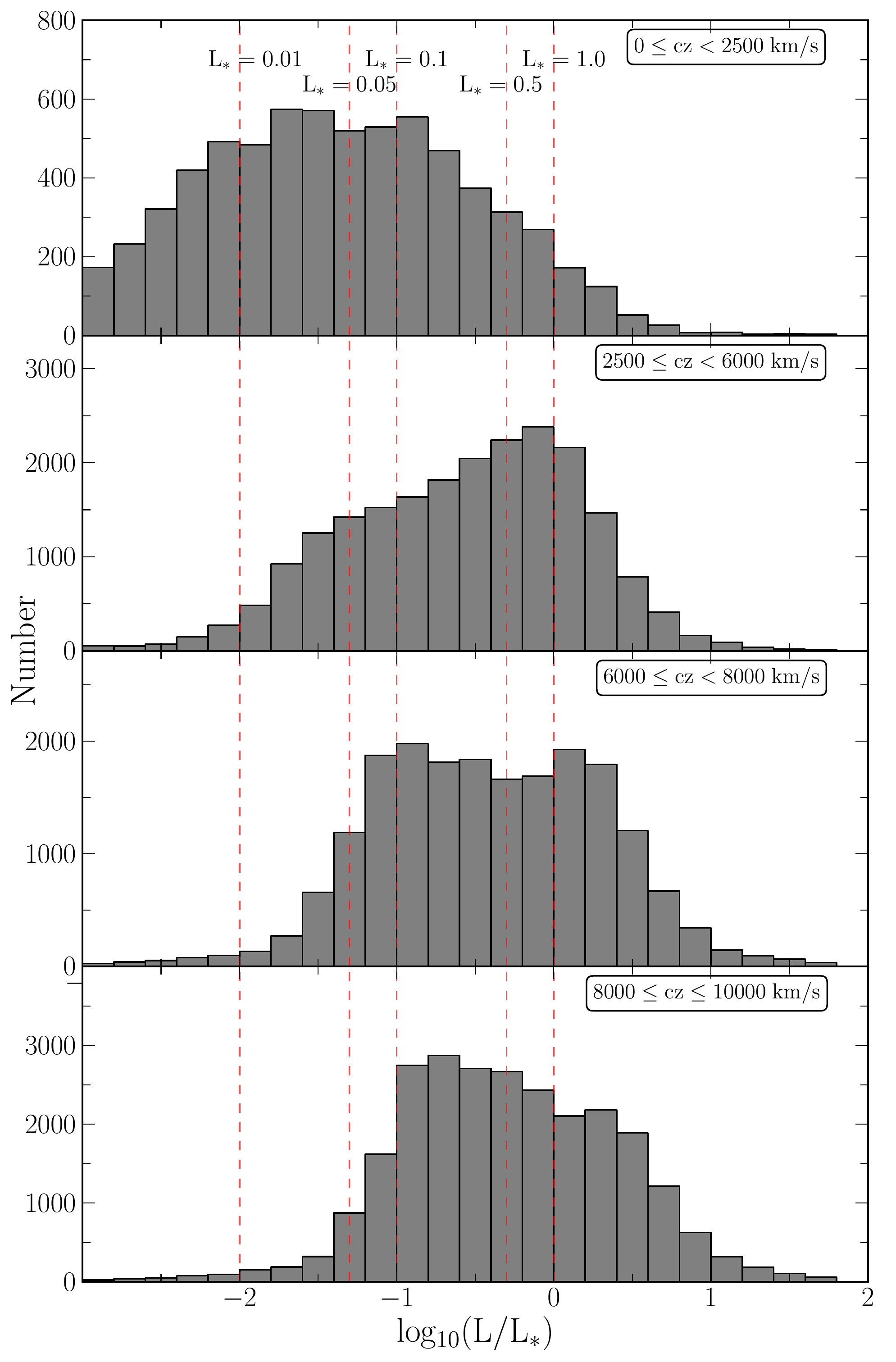}
        \caption{\small{Distribution of $L/L_{\**}$ values for all galaxies in the dataset. The red dashed vertical lines highlight 1, 0.5, 0.1, 0.05 and 0.01 $L_{\**}$.}}
        \label{completeness}
\end{figure} 

The galaxy data set contains over 108,000 entries, and includes data from SDSS, 2MASS, 2dF, 6dF, RC3, and many other, smaller surveys. Our criterion for including a galaxy in this data set is only an accurate, spectroscopic redshift that places the galaxy in the $cz \leq 10,000$ $\rm km\, s^{-1}$ velocity range. This restriction leads to a completeness limit of $B \lesssim 18.7$ mag, or $\sim0.2 L_*$, at $cz = 10,000$ $\rm km\, s^{-1}$, and progressively better towards lower velocities (see Figure \ref{completeness}). 

This limit will vary depending on which major surveys include a particular region of the sky. The major contributor is whether or not SDSS data are available, which begins around $cz = 5000$ $\rm km\, s^{-1}$. Figure \ref{completeness} is split into four velocity bins to illustrate this. Our data are complete down to $\sim0.1 L_*$ in the first bin, $0 \leq cz \leq 2500$ $\rm km\, s^{-1}$. At slightly higher velocities, $2500 \leq cz \leq 6000$ $\rm km\, s^{-1}$, the completeness falls to barely better than $\sim1.0 L_*$ as we move past the near and well studied galaxies, but have yet to reach the footprint of deep surveys. SDSS data become available in the last two bins, spanning $6000 \leq cz \leq 10,000$ $\rm km\, s^{-1}$. While the $8000 \leq cz \leq 10,000$ $\rm km\, s^{-1}$ region appears to reach the expected SDSS limits of $B \lesssim 18.7$ mag, or $\sim0.2 L_*$, the $6000 \leq cz \leq 8000$ $\rm km\, s^{-1}$ region instead appears to flatten below $\sim 1.0 L_*$. It is possible this is due to larger distance errors, since in this region redshift-independent distances are rare, and Hubble flow distances are still small enough to remain relatively uncertain.

Additionally, we have homogenized the galaxy data beyond the steps taken by NED by normalizing all measurements of galaxy inclination, position angle, and diameter to 2MASS $K$-band values. Most galaxies in NED have measures of inclination, position angle and diameter available in several different bands, so in order to make meaningful comparisons it is necessary to automatically choose one band for all measurements. We chose 2MASS values for this because it was an all-sky survey, and represents the largest fraction of available galaxy data. Physical galaxy diameters are derived from 2MASS $K_{s}$ ``total" angular diameter measurements and galaxy distances. 2MASS $K_{s}$ ``total" diameter estimates are surface brightness extrapolation measurements and are derived as 

\begin{equation}
	r_{\rm tot} = r' + a(\ln(148)^b,
\end{equation}

\noindent where $r_{tot}$ is defined as the point where the surface brightness extends to 5 disk scale lengths, $r'$ is the starting point radius ($>5" - 10"$ beyond the nucleus, or core influence), and $a$ and $b$ are Sersic exponential function scale length parameters ($f = f_0 \exp{(-r/a)}^{(1/b)}$; see Jarrett et al. 2003 for a full description). Approximately $50\%$ of all the galaxies have this 2MASS $K_{s}$ ``total" diameter. Of the remainder, $20\%$ have SDSS diameters, $27\%$ have no published diameter, and $3\%$ have diameters from other surveys. We convert values in these other bands to 2MASS $K_{s}$ ``total" diameters via a simple least squares linear fit when necessary.

We used $B$-band magnitudes to estimate each galaxy's luminosity in units of $L_{\**}$ as follows:

\begin{equation}
	\frac{L}{L_{\**}} = 10^{-0.4 (M_{\rm B} - M_{\rm B_{\**}})}.
\end{equation}

We adopt the CfA galaxy luminosity function by Marzke et al. (1994), which sets $B_{\**} $ = -19.57. Direct $B$-band measurements are available for $\sim 30\%$ of galaxies, and most of the rest have SDSS $g$ and $r$ magnitudes, which can be converted to $B$ via $B = g + 0.39 (g-r) + 0.21$ (Jester et al. 2005). Finally, we also compute an estimate of the virial radius of each galaxy as $\log R_{vir} = 0.69 \log D + 1.24$. This follows the parametrization of Stocke et al. (2013) relating a galaxy's luminosity to its virial radius, and the Wakker $\&$ Savage (2009) empirical relation between diameter and luminosity (see Wakker et al. 2015 and references therein for further details). Errors are propagated from the original published magnitude errors.

This homogeneous galaxy data table allows us to draw direct comparisons between the properties of the absorbers and the properties, separations, and environments of nearby galaxies with unprecedented completeness. The full dataset will be publicly released and discussed in further detail in a forthcoming paper (D. M. French $\&$ B. P. Wakker 2017, in preparation).

\subsection{Spectra}

This initial pilot study contains 33 sightlines to bright QSOs observed with COS. We chose sightlines by first sorting the galaxy data table described above by galaxy diameter. This sorted list is then correlated with the full list of publicly available sightlines, and only those systems with impact parameters less than 500 Mpc and galaxy diameters, $D$, greater than $25$ kpc, are kept. Finally, we select 33 sightlines with $S/N \geq 10$ from this list (see Table \ref{table1}).

All COS spectra for the target sightlines were obtained through the Barbara A. Mikulski Archive for Space Telescopes (MAST), and processed with CALCOS v3.0 or later. We combined individual exposures by the method of Wakker et al. (2015), which corrects the COS wavelength scale by cross-correlating all ISM and IGM lines in each exposure. This method addresses the up to $\pm40$ $\rm km\, s^{-1}$ misalignments produced by CALCOS, and produces a corrected error array based on Poisson noise, which better matches the measured errors than the errors delivered in the x1d files. We then combine multiple exposures by aligning Galactic absorption lines with 21 cm spectra, and adding up the total counts in each pixel before converting to flux using the original, average flux-count ratio at each wavelength.

Each absorption component is treated individually. Of our sample of 48, 13 are partially blended with another line (see Figure \ref{line} as an example - 12 of these 13 are likewise blended with another Ly$\alpha$ line), and 35 are distinct, such that the flux returns to the continuum level between the lines. The result of this method is that 14 galaxies are associated with multiple Ly$\alpha$ systems. Equivalent widths are measured by first performing a low-order (3rd order or lower) polynomial continuum fit in the line region. Then we integrate over the absorption velocity range, and calculate errors by the method of Wakker et al. (2003), which combines the random noise errors, the uncertainty of the continuum location, fixed-pattern noise, and the uncertainty in choosing the absorption velocity edges. Finally, a Gaussian is fit to the absorption profile to determine the line centroid.

\begin{table*}[ht]
\caption{\small{All Associated Systems Galaxy-absorber Systems}}
  \vspace{-20pt}
\begin{center}
\begin{tabular}{l c l l l l l l l l l l l l l}
 \hline \hline
 $\rm Target$ & $\rm Galaxy$ & $R_{vir}$         & $\rm L/L_{\**}$ & $v_{galaxy}$    &  $\rm Inc.$          &  $\rm Az.$ 	     & $\rho$		& $v_{Ly\alpha}$	& $W_{Ly\alpha}$   & $\Delta v$  	     & $\mathcal{L}$  \\ 
  	   	     &                        & \scriptsize (kpc) &          	      & \scriptsize (\kms) & \scriptsize (deg) & \scriptsize (deg) & \scriptsize (kpc) & \scriptsize (\kms) & \scriptsize (\kms) & \scriptsize (\kms) & 		\\
\scriptsize (1) & \scriptsize (2) & \scriptsize (3) & \scriptsize (4) & \scriptsize (5)      & \scriptsize (6)     & \scriptsize  (7)   & \scriptsize (8)    & \scriptsize (9)        & \scriptsize (10)     & \scriptsize (11)    & \scriptsize (12) \\ \hline \hline
1H0717+714  				&  UGC03804  					&  173  & 1.9 &  2887  	&  55  &  7  	&  207  &  2870  	&  343$\pm$6  		&  17  	&  0.24   \\
1H0717+714  				&  UGC03804  					&  173  & 1.9 &  2887  	&  55  &  7  	&  207  &  2956  	&  39$\pm$4  		&  -69  	&  0.21  \\
2dFGRS\_S393Z082  		&  NGC1097  					&  304  & 6.1 &  1271  	&  58  &  27  	&  112  &  1239  	&  570$\pm$21  	&  32  	&  1.9*  \\
H1101-232  				&  MCG-04-26-019  				&  173  & 1.1 &  3623  	&  68  &  26  	&  179  &  3580  	&  573$\pm$12  	&  43  	&  0.33  \\
HE0241-3043  				&  NGC1097  					&  304  & 6.1 &  1271  	&  58  &  77  	&  219  &  1221  	&  83$\pm$12  		&  50  	&  1.6*  \\
HE0241-3043  				&  NGC1097  					&  304  & 6.1 &  1271  	&  58  &  77  	&  219  &  1310  	&  184$\pm$15  	&  -39  	&  1.6*  \\
LBQS1230-0015  			&  NGC4517  					&  208  & 0.5 &  1128  	&  90  &  90  	&  110  &  1127  	&  473$\pm$16  	&  1  		&  1.6*  \\
MRC2251-178  			&  MCG-03-58-009  				&  319  & 2.3 &  9030  	&  61  &  39  	&  320  &  9051  	&  60$\pm$4  		&  -21  	&  1.4*  \\
Mrk1014  					&  NGC0768  					&  231  & 3.0 &  7021  	&  64  &  85  	&  486  &  7080  	&  117$\pm$11  	&  -59  	&  0.042*  \\
Mrk290  					&  NGC5987  					&  322  & 3.0 &  3010  	&  67  &  12  	&  486  &  3105  	&  511$\pm$5  		&  -95  	&  0.77*  \\
Mrk290  					&  NGC5987  					&  322  & 3.0 &  3010  	&  67  &  12  	&  486  &  3207  	&  319$\pm$4  		&  -197  	&  0.37*  \\
Mrk876  					&  UGC10294  					&  165  & 0.1 &  3504  	&  51  &  7  	&  274  &  3478  	&  280$\pm$3  		&  26  	&  0.063  \\
PG0003+158  				&  NGC7814  					&  171  & 1.2 &  1050  	&  68  &  47  	&  197  &  833  		&  131$\pm$15  	&  217  	&  0.081  \\
PG0832+251  				&  KUG0833+252  				&  165  & 0.7 &  6964  	&  62  &  55  	&  294  &  6980  	&  133$\pm$14  	&  -16  	&  0.041  \\
PG0832+251  				&  KUG0833+252  				&  165  & 0.7 &  6964  	&  62  &  55  	&  294  &  7201  	&  48$\pm$10  		&  -237  	&  0.01  \\
PG1001+054  				&  UGC05432  					&  164  & 1.3 &  3995  	&  36  &  78  	&  217  &  4092  	&  222$\pm$10  	&  -97  	&  0.14  \\
PG1302-102  				&  NGC4939  					&  235  & 4.4 &  3110  	&  48  &  61  	&  265  &  3448  	&  71$\pm$5  		&  -338  	&  0.05*  \\
RBS1768  				&  RFGC3781  					&  253  & 1.0 &  9162  	&  90  &  74  	&  464  &  9360  	&  364$\pm$4  		&  -198  	&  0.056*  \\
RBS1768  				&  RFGC3781  					&  253  & 1.0 &  9162  	&  90  &  74  	&  464  &  9434  	&  160$\pm$5  		&  -272  	&  0.024*  \\
RX J0714.5+7408  			&  UGC03717  					&  202  & 1.2 &  4188  	&  63  &  83  	&  271  &  4074  	&  58$\pm$7  		&  114  	&  0.13*  \\
RX J0714.5+7408  			&  UGC03717  					&  202  & 1.2 &  4188  	&  63  &  83  	&  271  &  4264  	&  410$\pm$9  		&  -76  	&  0.15*  \\
RX J1017.5+4702  			&  NGC3198  					&  191  & 1.3 &  663  	&  73  &  55  	&  378  &  629  		&  60$\pm$17  		&  34  	&  0.02  \\
RX J1117.6+5301  			&  NGC3631 					&  187  & 1.9 &  1156  	&  16  &  47  	&  198  &  1131  	&  356$\pm$20  	&  25  	&  0.32  \\
RX J1117.6+5301  			&  NGC3631 	 				&  187  & 1.9 &  1156  	&  16  &  47  	&  198  &  1259  	&  57$\pm$17  		&  -103  	&  0.25  \\
RX J1236.0+2641  			&  NGC4559  					&  165  & 0.7 &  807  	&  64  &  31  	&  188  &  795  		&  295$\pm$37  	&  12  	&  0.27  \\
RX J1236.0+2641  			&  NGC4565  					&  292  & 1.7 &  1230  	&  90  &  39  	&  159  &  1012  	&  337$\pm$32  	&  218  	&  0.54*  \\
RX J1236.0+2641  			&  NGC4565  					&  292  & 1.7 &  1230  	&  90  &  39  	&  159  &  1188  	&  288$\pm$24  	&  42  	&  1.7*  \\
RX J1330.8+3119  			&  UGC08492  					&  204  & 2.0 &  7414  	&  16  &  41  	&  335  &  7401  	&  330$\pm$15  	&  13  	&  0.081*  \\
RX J1356.4+2515  			&  CGCG132-055  				&  206  & 1.3 &  8671  	&  36  &  25  	&  190  &  8475  	&  126$\pm$18  	&  196  	&  0.35*  \\
RX J1503.2+6810  			&  CGCG318-012  				&  250  & 2.1 &  9765  	&  52  &  1  	&  325  &  10122  	&  44$\pm$14  		&  -357  	&  0.031*  \\
RX J1544.5+2827  			&  CGCG166-047  				&  175  & 1.8 &  9646  	&  43  &  61  	&  326  &  9642  	&  183$\pm$14  	&  4  		&  0.031  \\
RX J1544.5+2827  			&  CGCG166-047  				&  175  & 1.8 &  9646  	&  43  &  61  	&  326  &  9759  	&  169$\pm$12  	&  -113  	&  0.023  \\
RX J2043.1+0324  			&  NGC6954  					&  166  & 1.4 &  4067  	&  56  &  66  	&  301  &  4080  	&  82$\pm$10  		&  -13  	&  0.037  \\
RX J2139.7+0246  			&  UGC11785  					&  203  & 0.7 &  4074  	&  90  &  69  	&  108  &  4083  	&  490$\pm$7  		&  -9  	&  1.5  \\
RX J2139.7+0246  			&  UGC11785  					&  203  & 0.7 &  4074  	&  90  &  69  	&  108  &  4181  	&  529$\pm$7  		&  -107  	&  1.2*  \\
SBS0957+599  			&  MCG+10-14-058  				&  261  & 1.1 &  9501  	&  75  &  19  	&  206  &  9469  	&  78$\pm$12  		&  32  	&  1.4*  \\
SDSSJ021218.32-073719.8  	&  SDSSJ021315.79-			&  174  & 1.8 &  4800  	&  52  &  10  	&  268  &  4756  	&  528$\pm$15  	&  44  	&  0.09  \\
					  	& 			073942.7  		& 	    &	     &		  	&  	  &  	  	&	    &  		  	&  			  	&  	  	&  	  \\
SDSSJ021218.32-073719.8  	&  SDSSJ021315.79-		 	&  174  & 1.8 &  4800  	&  52  &  10  	&  268  &  4833  	&  500$\pm$17  	&  -33  	&  0.092  \\
					  	& 			073942.7  		& 	    &	     &		  	&  	  &  	  	&	    &  		  	&  			  	&  	  	&  	  \\
SDSSJ080838.80		  	&  UGC04239  					&  279  & 2.1 &  8763  	&  45  &  38  	&  378  &  8740  	&  883$\pm$24  	&  23  	&  0.87*  \\
		+051440.0	  	& 					  		& 	    &	     &		  	&  	  &  	  	&	    &  		  	&  			  	&  	  	&  	  \\
SDSSJ080838.80+051440.0  	&  UGC04239  					&  279  & 2.1 &  8763  	&  45  &  38  	&  378  &  8927  	&  130$\pm$19  	&  -164  	&  0.45*  \\
		+051440.0	  	& 					  		& 	    &	     &		  	&  	  &  	  	&	    &  		  	&  			  	&  	  	&  	  \\
SDSSJ091728.60		  	&  UGC04895  					&  204  & 2.1 &  7073  	&  61  &  32  	&  408  &  7141  	&  374$\pm$23  	&  -68  	&  0.022*  \\
		+271951.0	  	& 					  		& 	    &	     &		  	&  	  &  	  	&	    &  		  	&  			  	&  	  	&  	  \\
SDSSJ112224.10		  	&  NGC3640  					&  180  & 2.8 &  1251  	&  38  &  22  	&  139  &  1049  	&  288$\pm$30  	&  202  	&  0.4  \\
		+031802.0	  	& 					  		& 	    &	     &		  	&  	  &  	  	&	    &  		  	&  			  	&  	  	&  	  \\
SDSSJ112224.10+031802.0  	&  NGC3640  					&  180  & 2.8 &  1251  	&  38  &  22  	&  139  &  1264  	&  424$\pm$27  	&  -13  	&  1.1  \\
		+031802.0	  	& 					  		& 	    &	     &		  	&  	  &  	  	&	    &  		  	&  			  	&  	  	&  	  \\
SDSSJ130524.30		  	&  UGC08186  					&  268  & 1.1 &  7006  	&  82  &  14  	&  249  &  7039  	&  480$\pm$14  	&  -33  	&  1.3*  \\
		+035731.0	  	& 					  		& 	    &	     &		  	&  	  &  	  	&	    &  		  	&  			  	&  	  	&  	  \\
SDSSJ135726.27		  	&  NGC5364  					&  211  & 2.4 &  1241  	&  57  &  84  	&  183  &  1124  	&  85$\pm$11  		&  117  	&  0.74*  \\
		+043541.4	  	& 					  		& 	    &	     &		  	&  	  &  	  	&	    &  		  	&  			  	&  	  	&  	  \\
SDSSJ135726.27+043541.4  	&  NGC5364  					&  211  & 2.4 &  1241  	&  57  &  84  	&  183  &  1296  	&  98$\pm$9  		&  -55  	&  0.97*  \\
		+043541.4	  	& 					  		& 	    &	     &		  	&  	  &  	  	&	    &  		  	&  			  	&  	  	&  	  \\
SDSSJ140428.30		  	&  KUG1402+341  				&  204  & 1.1 &  7919  	&  72  &  63  	&  118  &  7884  	&  889$\pm$28  	&  35  	&  1.4  \\
		+335342.0	  	& 					  		& 	    &	     &		  	&  	  &  	  	&	    &  		  	&  			  	&  	  	&  	  \\
TON1009  				&  NGC2770  					&  204  & 1.9 &  1947  	&  87  &  41  	&  274  &  1961  	&  350$\pm$21 	&  -14  	&  0.19*  \\

 \\
\hline
\end{tabular}

\footnotesize \raggedright \textbf{Note.} The largest $\mathcal{L}$ value is given, with a (\**) indicating that this corresponds to $\mathcal{L}_{\rm D^{1.5}}$, otherwise the quoted $\mathcal{L}$ was computed with $R_{vir}$.
\end{center}
  \label{target_table}
\end{table*}

\section{RESULTS}

We have identified 48 Ly$\alpha$ absorption lines in the spectra of our initial 33 QSO sample, each of which has been associated with a single nearby galaxy of diameter $D\geq25$ kpc. Each absorption component is treated individually, resulting in several cases where multiple absorbers are associated with the same galaxy. In order to be considered for a pairing, a galaxy and absorption feature must appear within 400 $\rm km\, s^{-1}$ in velocity and 500 kpc in physical impact parameter from each other. When multiple galaxies pass these criteria for a particular line, we are left with two options: (1) one galaxy is obviously far larger and closer in physical and velocity space to the sightline, and may have several satellite galaxies; or (2) there are multiple galaxies near the absorber, making any association ambiguous; we do not include these cases in the further analysis.

\begin{figure}[t!]
\centering
  \subfigure[]{\includegraphics[width=0.87\linewidth]{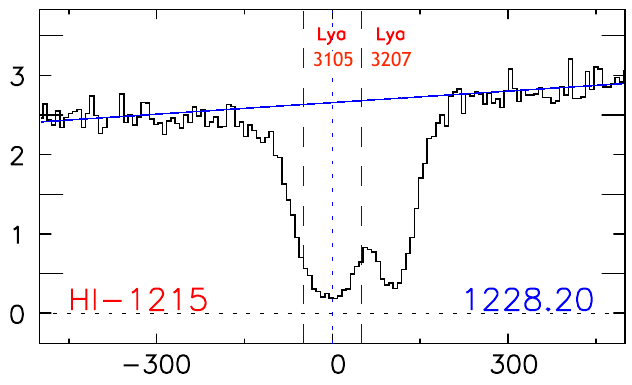}\label{line}}
  \subfigure[]{\includegraphics[width=1.\linewidth]{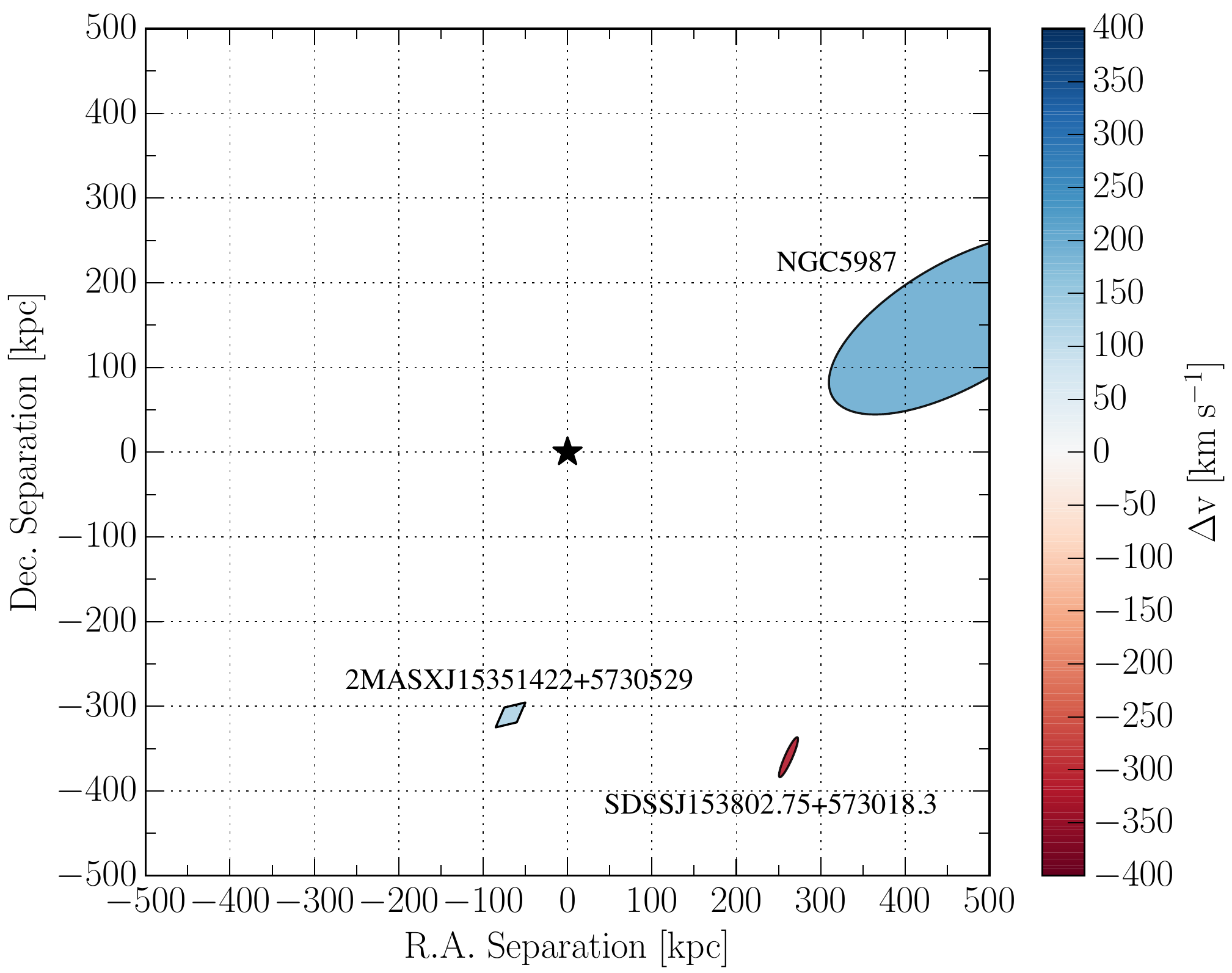}\label{impactmap}}
  \caption{\small{(a) An example of 2 Ly$\alpha$ lines found in the Mrk290 sightline at 3105 and 3207 . (b) A map of \textit{all} galaxies within a 500 kpc impact parameter of target Mrk290 sightline and with velocity ($cz$) within 400 $\rm km\, s^{-1}$ of absorption detected at 3207 $\rm km\, s^{-1}$ (central black star). The galaxy NGC5987 ($v=3010$ $\rm km\, s^{-1}$, inclination = $65^{\circ}$) has been paired with the Ly$\alpha$ absorption features at $v=3105, 3207$ $\rm km\, s^{-1}$ because it is the largest and closest galaxy in both physical and velocity space to the absorption feature.}}
\vspace{5pt}
\end{figure}

\begin{figure*}[t]
\centering
\subfigure[]{\label{ew_vs_impact}\includegraphics[width=0.49\textwidth]{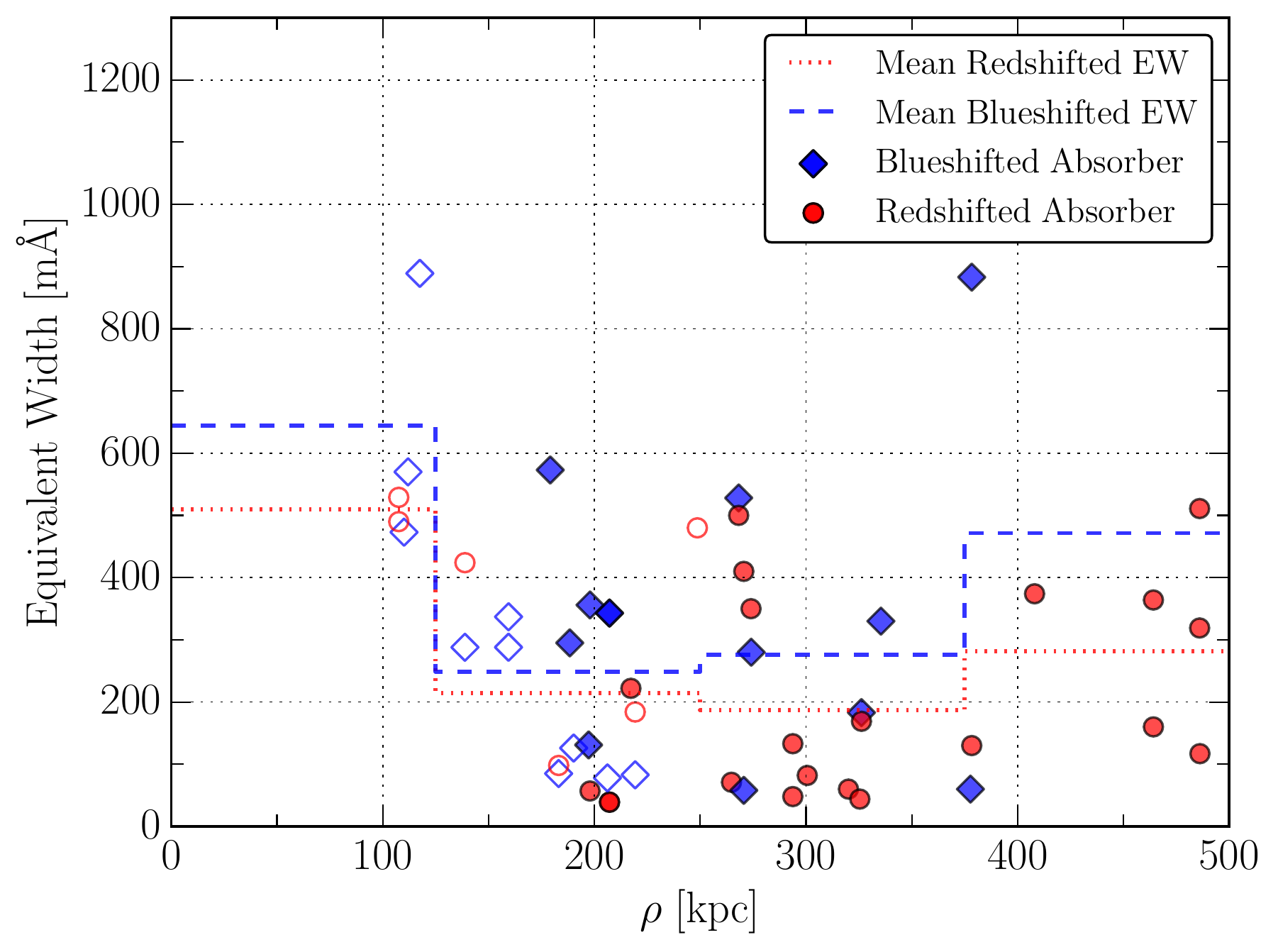}}
\subfigure[]{\label{ew_vs_impact_vir}\includegraphics[width=0.49\textwidth]{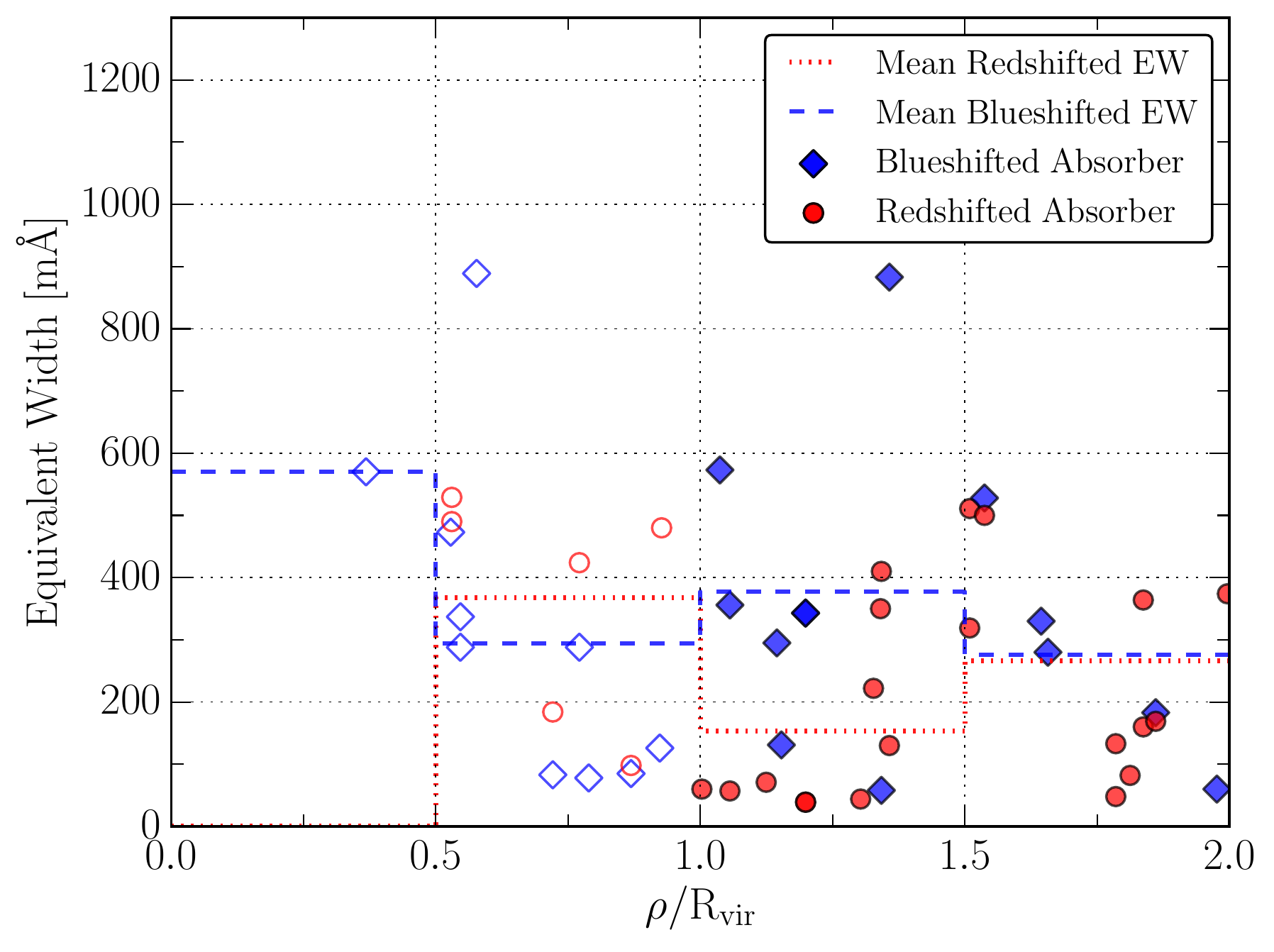}}
\caption{\small{(a) Equivalent width of each absorber as a function of impact parameter $\rho$. (b) Equivalent width as a function of $\rho/R_{vir}$. The anti-correlation is strongest when scaling $\rho$ by the galaxy virial radius. Absorbers are separated into redshifted and blueshifted samples based on $\Delta v$. Bins of mean $EW$ are overplotted in red dotted, and blue dashed lines for their respective samples. Open symbols correspond to systems with $\rho \leq R_{vir}$.}}
\vspace{5pt}
\end{figure*}

To facilitate this decision, we calculate the likelihood, $\mathcal{L}$, of every possible galaxy-absorber pairing as follows:

\begin{equation}
	\mathcal{L} = A e^{-(\frac{\rho}{R_{\rm eff}})^2} e^{-(\frac{\Delta v}{v_{\rm norm}})^2}.
\end{equation}

\noindent Here $\rho$ is the physical impact parameter, $\Delta v$ is the velocity difference between the absorber and the galaxy ($\Delta v = v_{galaxy} - v_{absorber}$), $v_{norm}$ is the velocity normalization constant, and $A$ is a factor included to increase the likelihood in the case that $\rho \leq R_{eff}$ (in which case $A = 2$, otherwise $A = 1$). Many similar studies and simulations (e.g., Wakker $\&$ Savage 2009; Liang $\&$ Chen 2014; Mathes et al. 2014) suggest that Ly$\alpha$ absorbers lie within 400 $\rm km\, s^{-1}$ of their associated galaxies, so throughout this paper we adopt a halfway point of $v_{norm}$ = 200 $\rm km\, s^{-1}$. Future work will explore the result of varying this normalization parameter and making refinements such as, e.g., relating $v_{norm}$ to the galaxy's rotation velocity.

We compute $\mathcal{L}$ for two different values of $R_{eff}$: $R_{vir}$, the virial radius of the galaxy, and $D^{1.5}$, the major diameter of the galaxy to the power of 1.5. $\mathcal{L}$ computed with $R_{vir}$ is liable to select satellite galaxies instead of the larger hosts, so including a version with $D^{1.5}$ serves as a two-tiered selection system. An absorber separated by 200 $\rm km\, s^{-1}$ in velocity and 1$R_{vir}$ in impact parameter from a $D=30$ kpc galaxy would have $\mathcal{L}_{R_{vir}} = 0.27$ and $\mathcal{L}_{D^{1.5}} = 0.11$. In order for an absorber to be marked as ``associated" with a particular galaxy, we require that its $\mathcal{L}$ must be a factor of 5 larger than the next best possible association, and $\mathcal{L} \ge 0.01$ for at least one of $\mathcal{L}_{R_{vir}}$ or $\mathcal{L}_{D^{1.5}}$. We visually inspect systems with only one $\mathcal{L}$ meeting these criteria, and decide to reject or include it based on the complexity of the nearby galaxy environment. In Table \ref{target_table} we quote only the largest value of $\mathcal{L}$, and use an asterisk to denote when this corresponds to $\mathcal{L}_{D^{1.5}}$.

Figures \ref{line} and \ref{impactmap} show an example of a Ly$\alpha$ absorption line with a map of its galaxy environment, showing an unambiguous pairing between the absorption features at $3105, 3207$ $\rm km\, s^{-1}$ toward Mrk290 and galaxy NGC5987 ($\mathcal{L}^* = 0.37$). All analysis that follows concerns similarly ``associated" systems.

Additionally, we split the absorber-galaxy catalog based on the velocity difference of the two, $\Delta v$. With this scheme, we refer to an absorber with a lower velocity than the associated galaxy as blueshifted, while an absorber with a higher velocity is referred to as redshifted. The rest of the results will be analyzed based upon this splitting. In all figures blue and redshifted absorbers are represented as blue diamonds and red circles, respectively, and red diamonds correspond to systems where both redshifted and blueshifted absorbers are detected. We use open symbols for systems with $\rho \leq R_{vir}$.

\begin{figure*}[ht]
\centering
\subfigure[]{\label{w_vir}\includegraphics[width=0.49\textwidth]{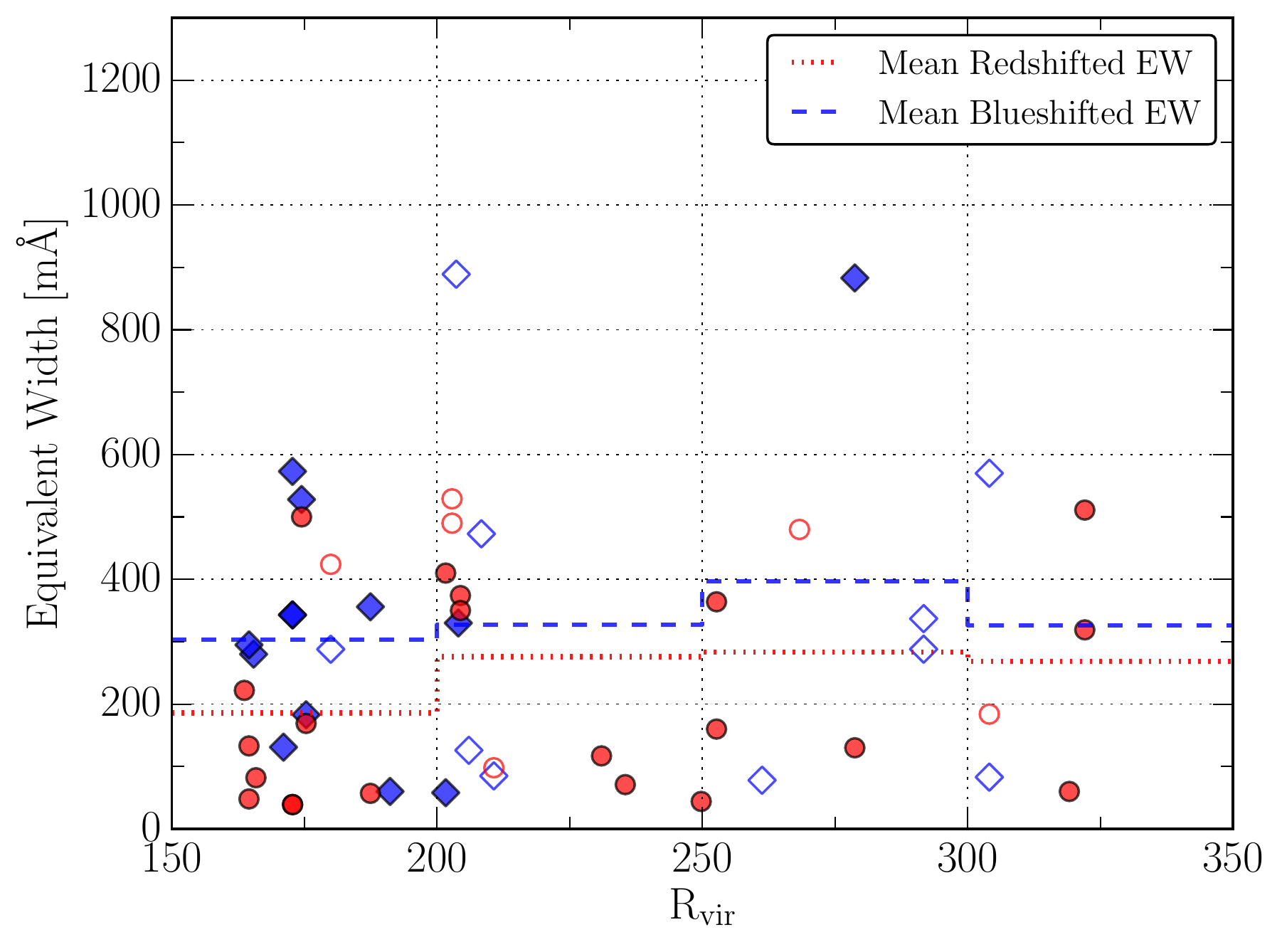}}
\subfigure[]{\label{impact_vir}\includegraphics[width=0.49\textwidth]{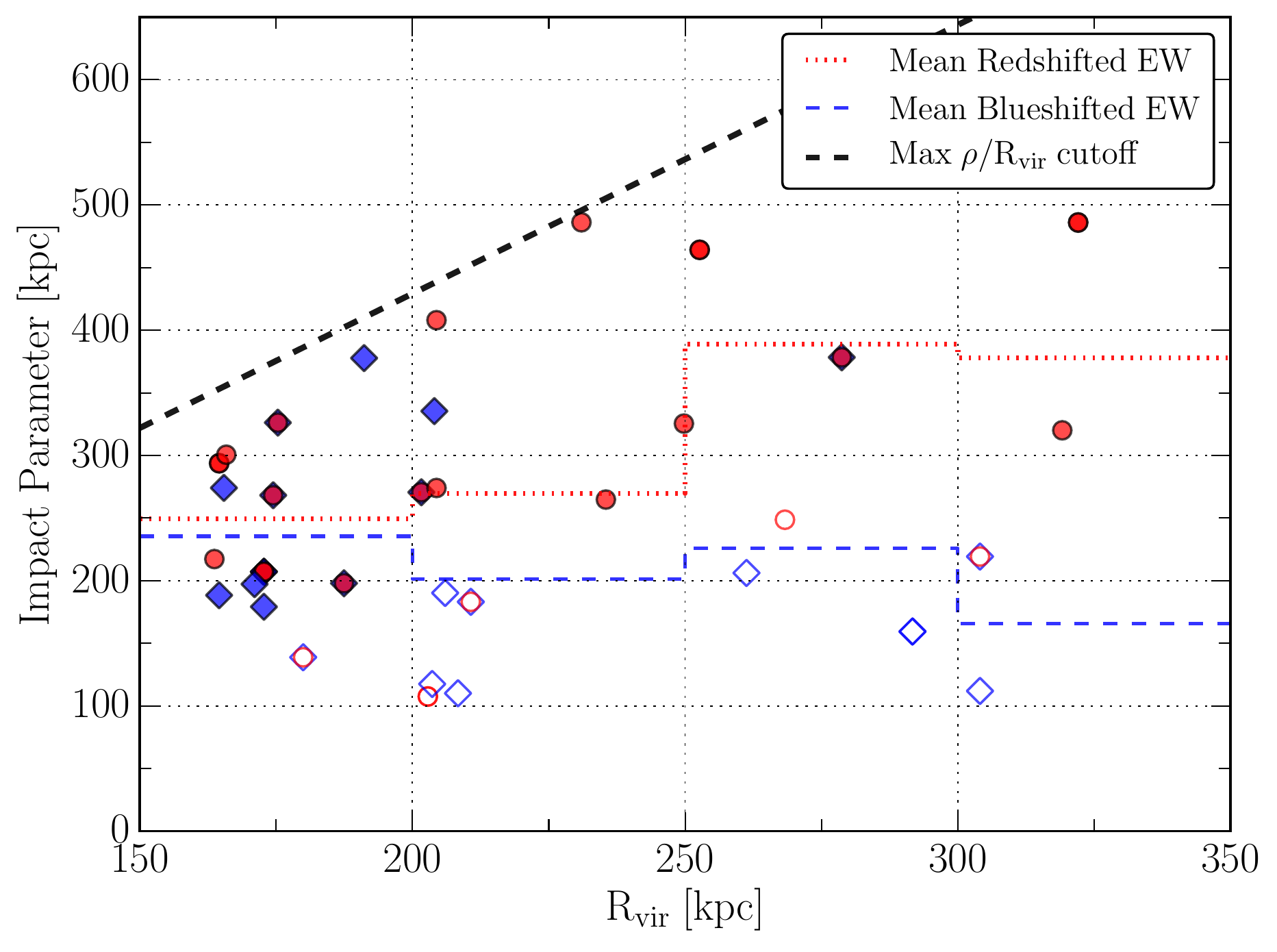}}
\caption{\small{(a) Equivalent width of each absorber as a function of the virial radius of the associated galaxy. (b) Impact parameter to each absorber as a function of the virial radius of the associated galaxy. The black dashed line indicates the cutoff at $\rho/R_{vir} =2.14$ imposed by our $\mathcal{L}$ limit.} In each, the blue dashed and red dotted lines show the average $EW$ in 50 kpc bins of impact parameter for the blueshifted and redshifted absorbers, respectively. Open symbols correspond to systems with $\rho \leq R_{vir}$.}
\vspace{5pt}
\end{figure*}

\vspace{10pt}

\subsection{EW-$\rho$ Anti-correlation}
\label{ew}

Numerous previous studies have suggested that Ly$\alpha$ equivalent width ($EW$) is anti-correlated with impact parameter ($\rho$) to the nearest galaxy. We find a weak anti-correlation, as shown in Figure \ref{ew_vs_impact}. However, as Churchill et al. (2013a) also found with Mg \,{\sc ii} absorption, we find a stronger anti-correlation when we normalize $\rho$ by $R_{vir}$. Figure \ref{ew_vs_impact_vir} shows this expected anti-correlation when plotting $EW$ versus $\rho/R_{vir}$. A possible explanation for this trend is that larger galaxies host larger, more physically extended CGM halos. We would thus expect the absorber $EW$ to also correlate positively with $R_{vir}$. Figure \ref{w_vir} shows $EW$ as a function of $R_{vir}$, with the blue-dashed and red-dotted lines showing the average $EW$ in bins of 50 kpc of $R_{vir}$, showing little evidence of a correlation. However, by similarly plotting $\rho$ as a function of $R_{vir}$, we instead find some evidence that absorbers around larger galaxies tend to be found at higher impact parameters. While we expect the upper-left quadrant of this figure to be sparsely populated (our likelihood-based method would tend not to choose small galaxies at large impact parameters), it is unclear to us why the lower right quadrant (large galaxies with absorbers at low impact parameter) is also sparsely populated. The full-sized sample at the completion of our study should provide a clearer picture.

\begin{figure}[h!]
        \centering
        \includegraphics[width=0.49\textwidth]{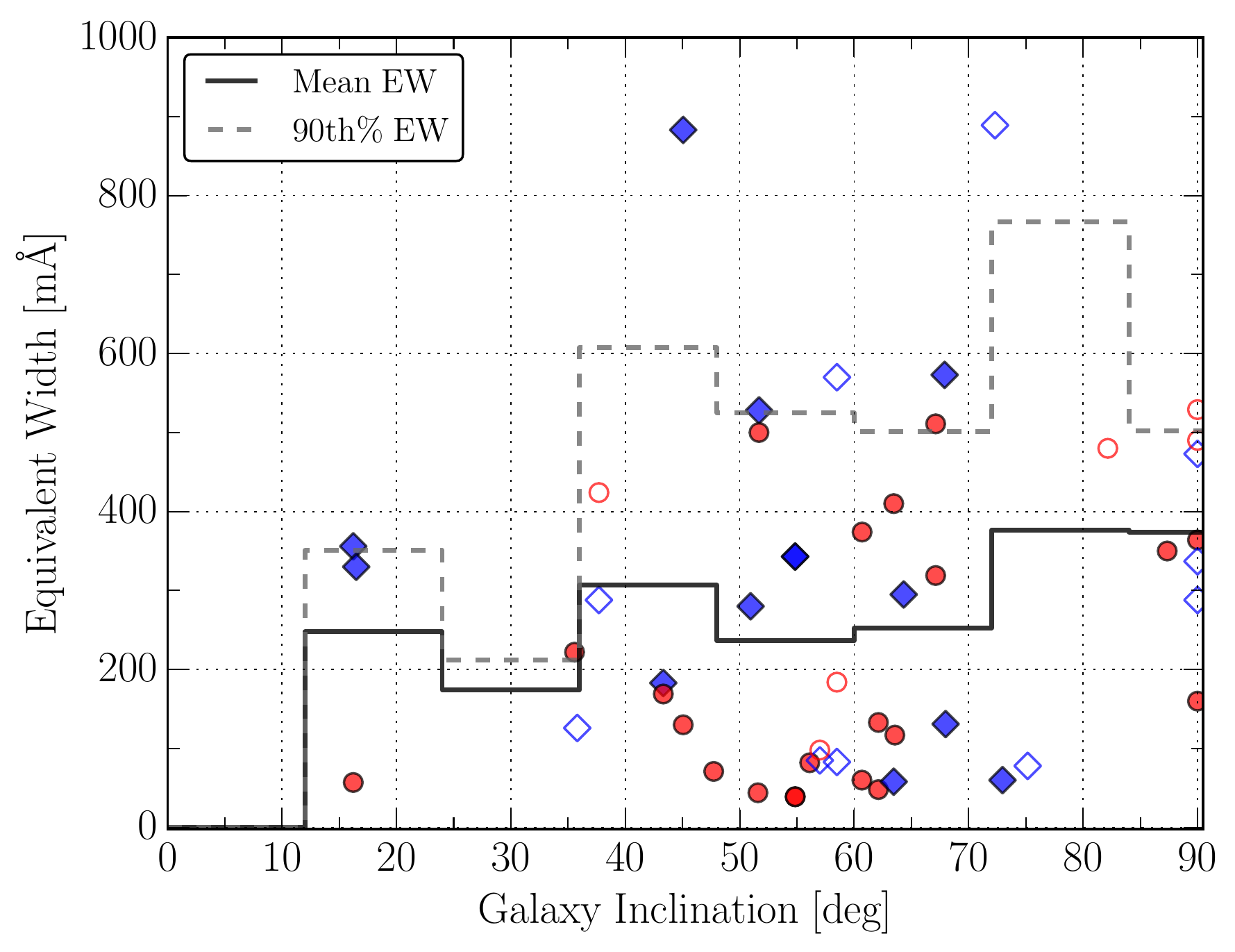}
        \caption{\small{Equivalent width of each absorber as a function of the inclination angle of the associated galaxy. The black and dashed gray lines show the mean and 90th percentile $EW$ of all absorbers in bins of $15^{\circ}$. Open symbols correspond to systems with $\rho \leq R_{vir}$.}}
        \label{ew_vs_inclination}
        \vspace{2pt}
\end{figure}

\begin{figure}[h!]
        \centering
        \includegraphics[width=0.49\textwidth]{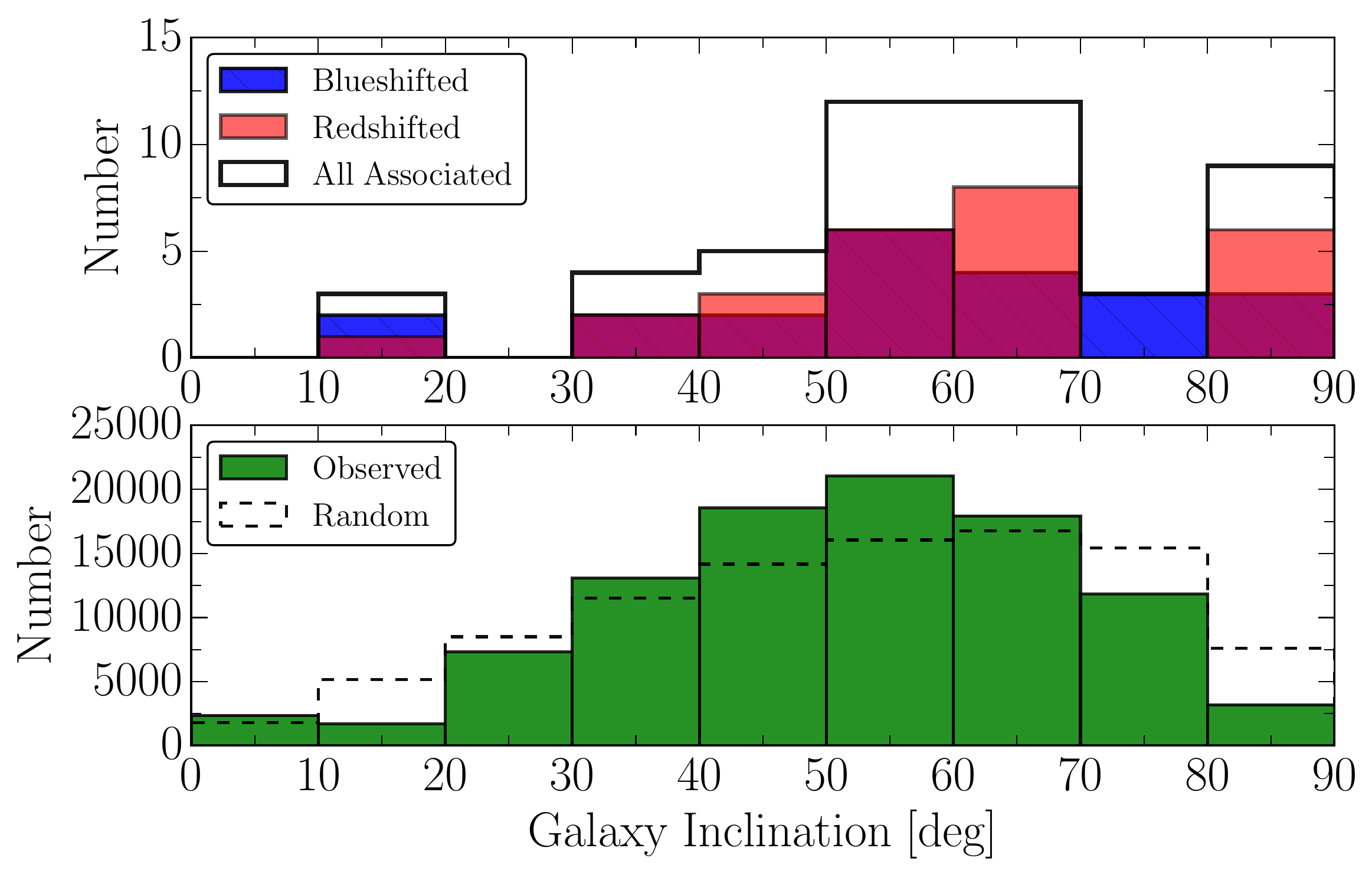}
        \caption{\small{Top: distribution of inclinations for all associated galaxies, split into redshifted and blueshifted sets. Bottom: distribution of inclinations of all observed galaxies in the $cz \leq 10,000$ $\rm km\, s^{-1}$ redshift range. The dashed line shows the inclination distribution for a truly random sample (i.e., no observational biases).}}
        \label{hist_inc}
        \vspace{2pt}
\end{figure}

\begin{figure}[ht!]
        \centering
        \includegraphics[width=0.49\textwidth]{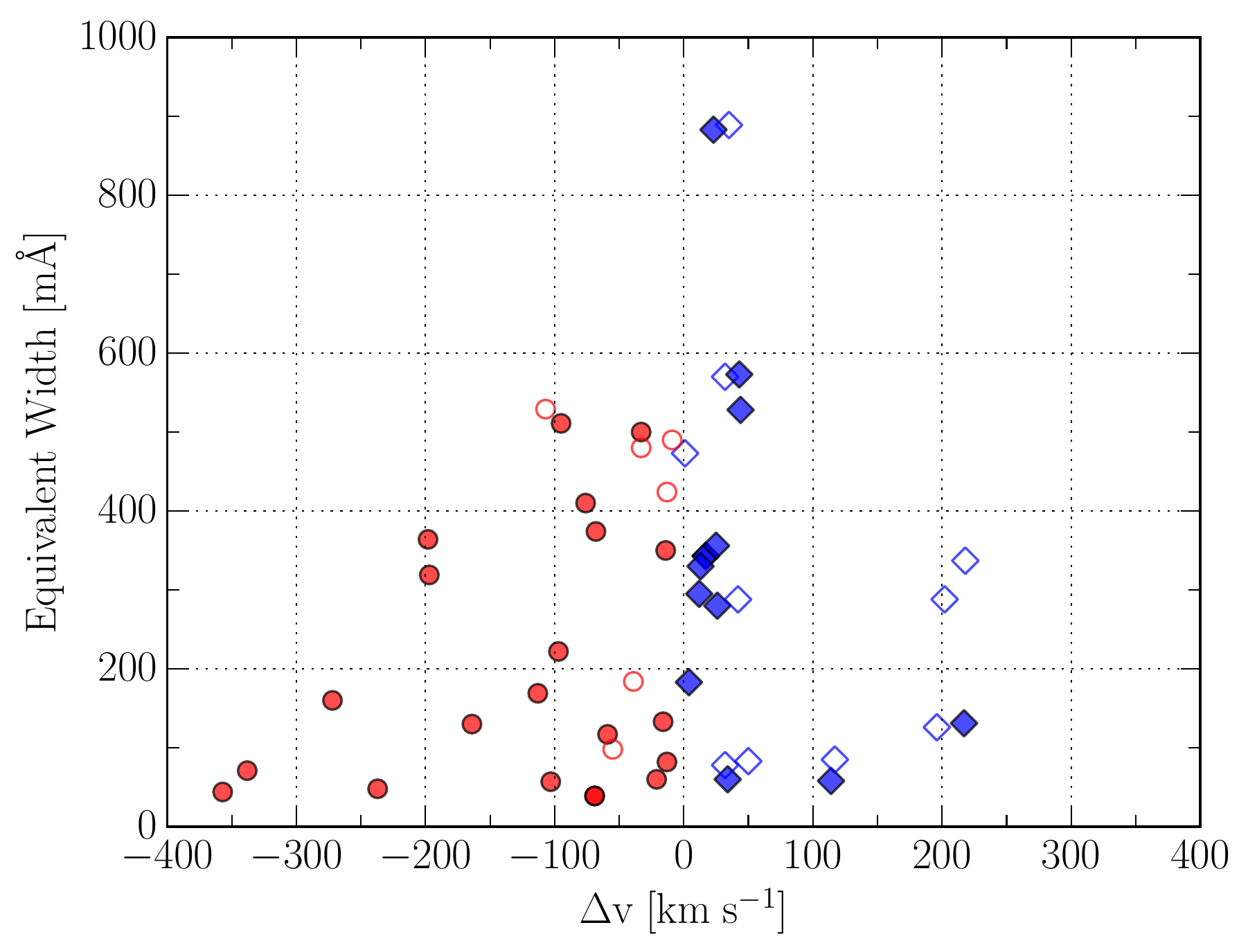}
        \caption{\small{Equivalent width as a function of the velocity separation between the galaxy and absorption line. Open symbols correspond to systems with $\rho \leq R_{vir}$.}}
        \label{W_veldif}
        \vspace{5pt}
\end{figure} 

\vspace{10pt}

\subsection{Inclination}
\label{inclination}

In this section we examine the inclinations of the associated galaxies compared to the distributions of absorbers. We correct for the finite thickness of galaxies, which causes $b/a$ to deviate from $\cos(i)$ at high inclinations, by computing galaxy inclinations with the following formula from Heidmann et al. (1972a):

\begin{equation}
	\cos(i) = \sqrt{\frac{q^2 - q_0^2}{1 - q_0^2}},
	\label{incEq}
\end{equation}

\noindent where $q = b/a$, the ratio of the minor to major axis, and $q_0$ is the intrinsic axis ratio, set to $q_0 = 0.2$ for all galaxies (e.g., Jones, Davies, and Trewhella 1996). Of the 48 total absorbers, 6 are associated with E or S0 type galaxies, but we have chosen to keep the value of $q_0 = 0.2$ uniform throughout. The calculated values of $\cos(i)$ that we use for these galaxies are thus conservative underestimates of their true inclinations.

Figure \ref{ew_vs_inclination} shows red and blueshifted absorbers' $EW$ plotted against the inclinations of their associated galaxies. We note that there is a clear excess of absorbers near galaxies of high inclination, with $77\%$ of redshifted and $73\%$ of blueshifted absorbers being associated with galaxies of $i \geq$ 50 deg, and only 3 absorbers being associated with a galaxy of $i<35$. The black and grey-dashed lines show mean and 90th percentile histograms, respectively, in bins of 12 deg. There does not appear to be much evolution of $EW$ across galaxy inclination, although a slight increase of mean $EW$ is possibly present toward higher inclinations.

In total, $75\%$ of absorbers are associated with highly inclined galaxies ($i \geq$ 50 deg). Only $56\%$ of all galaxies in the survey volume are highly inclined, indicating a preference for detecting absorption around inclined galaxies. Figure \ref{hist_inc} shows the distribution of galaxy inclinations for both the red and blue-shifted associated galaxies and all galaxies within the survey volume. We tested the difference between the full distribution of inclination angles for all galaxies in our survey volume and the distribution for all associated galaxies (red + blue-shifted absorbers) using the Anderson-Darling (AD) statistical distribution test, yielding a $p$-value of $AD_{p} = 0.00037$. Thus at a $99.96\%$ confidence level ($\sim 3.6 \sigma$ for a normal distribution) the inclinations of our associated galaxies are not sampled from the average distribution of observed inclinations. Hence, we take this to mean that the shape of the CGM of these galaxies is not perfectly spheroidal. 

It is worth noting here that the observed distribution of galaxy inclinations is \emph{not} flat, as one might naively expect. The dashed line in Figure \ref{hist_inc} shows the distribution of observable inclinations for a random, uniform sample (i.e., a uniform distribution of $q=b/a$ values between 0.2 and 1.0). There could be a number of effects contributing to the difference between this expected distribution and the observed (shown in green). If our sample is magnitude-limited and we assume galaxies are mostly optically thin but with a very-thin optically thick component (e.g., a dust lane), then mostly face-on and mostly edge-on galaxies would be underrepresented due to surface brightness and dust obscuration effects (e.g., see Jones, Davies and Trewhella 1996). It is possible that a similar effect is also responsible for the overabundance of Ly$\alpha$ detections around highly inclined galaxies. If we assume a disky or oblate spheroid halo shape and a covering fraction below unity for the CGM, the probability of encountering a cloud near an inclined galaxy would increase due to the increased path-length through the halo. We will produce a model to test this and other possible explanations in Paper II, when we have the much larger dataset available.

\vspace{10pt}

\subsection{Velocity Difference \rm($\Delta v$\rm)}
\label{veldiff}

We find evidence for an anti-correlation between absorber $EW$ and the velocity difference between the galaxy and the associated absorption, $\Delta v$. The mean and maximal $EW$ of absorption increases with decreasing $\Delta v$ (see Figure \ref{W_veldif}). In total, 32/48 ($67\%$) of absorbers are found within $\pm100$ $\rm km\, s^{-1}$. This $\pm100$ $\rm km\, s^{-1}$ threshold also applies to absorber $EW$, with only 1 absorber of $EW \geq 400$ found with $\Delta v > 100$ $\rm km\, s^{-1}$. 

Blueshifted absorbers are on average closer both in velocity and impact parameter to their associated galaxy, with $\overline{\Delta v}_{blue} = 68\pm16$ $\rm km\, s^{-1}$ and $\overline{\rho_{blue}} = 218\pm17$ kpc,  compared to $\overline{\Delta v}_{red}=-108\pm20$ $\rm km\, s^{-1}$ and $\overline{\rho_{red}} = 298\pm23$ kpc for the redshifted sample. Correspondingly, blueshifted absorbers have a slightly higher average equivalent width, $\overline{EW}_{blue}=329\pm52$ $\rm m\AA$ compared to $\overline{EW}_{red}=245\pm34$ $\rm m\AA$.

Additionally, of the 48 associated absorbers, 29 are matched with the same galaxy as another absorber (for a total of 14 unique galaxies in this subset). All but one of these cases involve two absorbers in the same sightline yet separated in velocity around a galaxy. Of these, 23 out of 29 are oriented such that the higher $EW$ absorber has the smaller $\Delta v$, and the 6 others are close in either velocity or $EW$. The one galaxy with three associated absorbers, NGC1097, shows this trend across two sightlines as well, with absorbers at $\Delta v = 32$ $\rm km\, s^{-1}$ and $EW = 570$ $\rm m\AA$ toward 2dFGRS\_S393Z082, and $\Delta v = -39$ $\rm km\, s^{-1}$ and $EW = 184$ $\rm m\AA$ and $\Delta v = 50$ $\rm km\, s^{-1}$ and $EW = 83$ $\rm m\AA$ toward HE0241-3043.

This result is the opposite of what we might expect selection effects associated with our likelihood method to produce. Because $\mathcal{L}$ is small for both high $\Delta v$ and high $\rho / R_{vir}$, there should be mostly low $\rho / R_{vir}$ systems at high $\Delta v$. Low $\rho / R_{vir}$ systems should also have higher $EW$ on average, as evidenced in the $EW-\rho$ anti-correlation discussed above. Figure \ref{W_veldif} shows the opposite, however, with only low-$EW$ systems at high $\Delta v$. It must therefore be the case that $EW$ tends to anti-correlate with both $\Delta v$ \textit{and} $\rho / R_{vir}$. Disentangling the relative strengths of the correlations between $EW$ and $\rho$, $R_{vir}$, and $\Delta v$ will require a larger data set, and thus we defer this discussion to our forthcoming Paper II of this series.

\vspace{10pt}

\subsection{Azimuth}
\label{azimuth}

\begin{figure}[t!]
        \centering
        \includegraphics[width=0.49\textwidth]{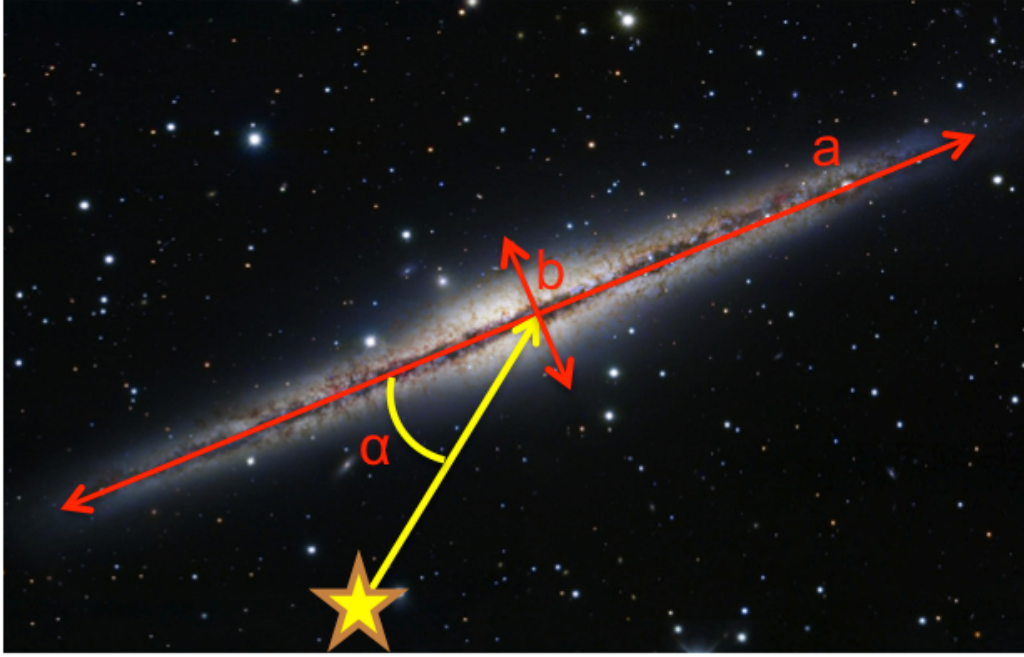}
        \caption{\small{Azimuth is the angle, $\alpha$, between the major axis of the galaxy, $a$, and a vector extending from the QSO target to the galaxy center. Image of NGC891 credit: composite Image Data - Subaru Telescope (NAOJ), Hubble Legacy Archive, Michael Joner, David Laney (West Mountain Observatory, BYU); Processing - Robert Gendler.}}
        \label{azimuth_illustration}
        \vspace{5pt}
\end{figure} 

\begin{figure}[t!]
        \centering
        \includegraphics[width=0.49\textwidth]{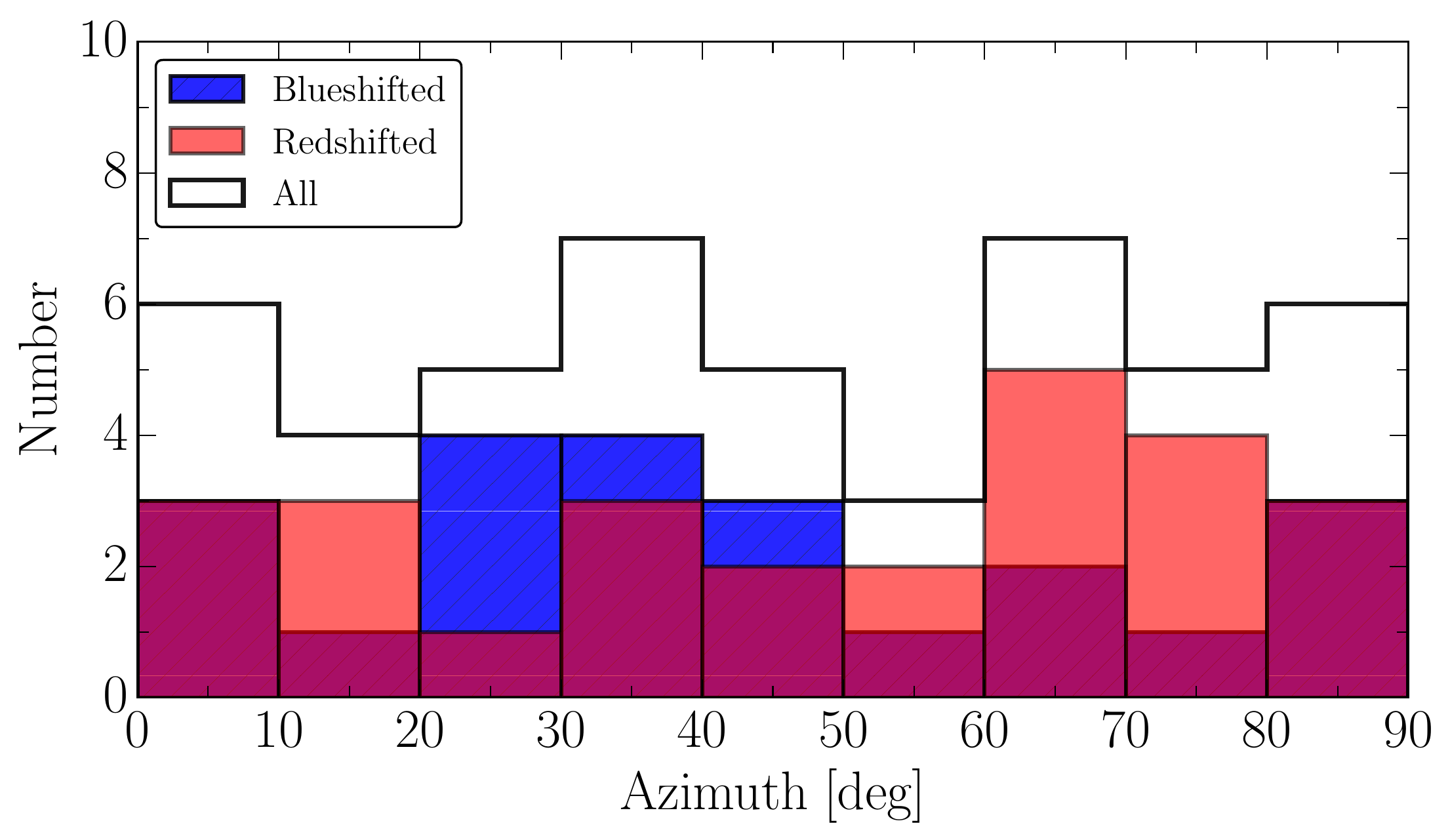}
        \caption{\small{The distributions of azimuth angles for red and blue-shifted samples, with the combined sample plotted in black. Azimuth $= 0$ corresponds to absorption detected along the projected major axis of the galaxy, and Azimuth $= 90$ is along the minor axis.}}
        \label{azimuth_dist}
\end{figure}

In this section we examine properties of absorbers as a function of their azimuthal angle with respect to their associated galaxy. Azimuth is defined as the angle between the major axis of a galaxy and the vector connecting the absorption feature and the midpoint of the galaxy plane. Figure \ref{azimuth_illustration} illustrates this. 

\begin{figure*}[ht!]
        \centering
        \includegraphics[width=0.98\textwidth]{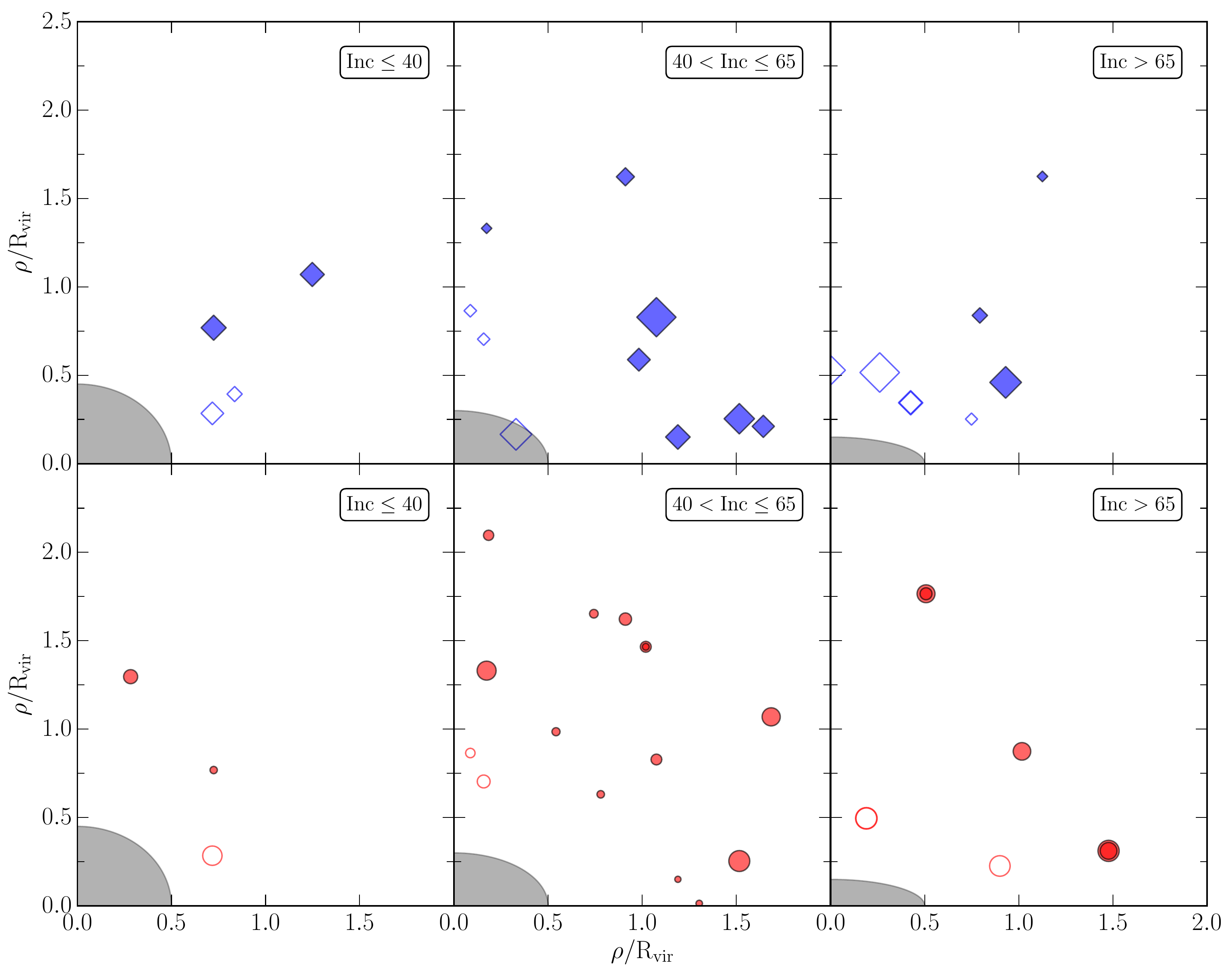}
        \caption{\small{A map of where each absorber was detected with respect to the associated galaxies, separated into three bins of inclination (illustrated by the gray ellipse in the bottom left corner of each plot). Left: absorbers associated with galaxies of inclination $0 \leq Inc \le 40$; center: $40 \leq Inc \le 65$; and right: $65 \leq Inc$. Blueshifted and redshifted absorbers are separated into the top (diamonds) and bottom (circles) panels, and the marker size is scaled with $EW$. The sightlines containing multiple absorbers associated with the same galaxy can be identified by their darker colors. Open symbols correspond to systems with $\rho \leq R_{vir}$.}}
        \label{azimuthMap}
\end{figure*}

The mean azimuth angle for blueshifted absorbers is $43\pm5^{\circ}$, and $49\pm5^{\circ}$ for redshifted absorbers. Figure \ref{azimuth_dist} shows the distribution of azimuth angles for both red and blue-shifted absorbers. Unlike the findings of Kacprzak et al. (2011b, 2012a), who find a bimodal distribution of Mg\,{\sc ii} absorbers around galaxies, our distributions of Ly$\alpha$ absorbers are generally consistent with a flat, or random distribution. There is possibly a slight overabundance of redshifted absorbers around $0^{\circ}$ (minor axis) and blueshifted absorbers between 20 and $50^{\circ}$ (just off major axis), but we cannot assign this observation much significance yet, given the small sample size. We additionally find no significant correlation between azimuth angle and $EW$ or $\Delta v$. See Figure \ref{azimuthMap} for a map of the locations of absorbers relative to their associated galaxies, split between redshifted and blueshifted absorbers and into three bins of inclination.

These contrasting results may indicate a genuine difference between the properties of Ly$\alpha$ and metal lines, since the distributions of the latter are thought to be influenced by outflows, which may be focused along the minor axis.

\section{DISCUSSION}

In this paper we have chosen to separate absorption systems into their individual components, whereas many comparable CGM studies have instead chosen to group together absorption components within some velocity window. While making the assumption that absorbers within some velocity interval are physically linked certainly has merit, it is also possible that absorption components close in velocity may in fact be physically distinct (see, e.g., Churchill et al. 2015). Indeed, in Section \ref{veldiff} we identified several systems where the $EW$ of components of a possible Ly$\alpha$ system individually anti-correlated with $\Delta v$. We will thus explore both methods in Paper II, where the much larger sample size will allow for a meaningful comparison.\\

Restricting ourselves to low-redshift systems has several benefits and consequences. We are able to extend our search for associated galaxies to larger impact parameters than many related, higher-redshift CGM studies, because of the availability of galaxy data. However, due to the observed anti-correlation between $EW$ and impact parameter, this results in a larger fraction of low-$EW$ and low-column density absorbers in our sample. Hence, we are likely tracing a region of the CGM not entirely analogous to that traced by, e.g., Kacprzak et al. (2011b), Mathes et al. (2014), and Borthakur et al. (2015).

Studies focusing on metal lines (e.g., Mg\,{\sc ii} and O\,{\sc vi}) are generally associated with high column density Lyman Limit Systems (LLS), which, again, tend to originate closer to their host galaxies. Most of the Ly$\alpha$ absorbers in this work are low column density (generally log$~N$(H\,{\sc i}) $\leq$ 14), and originate near or beyond 1 $R_{vir}$. At these distances we may actually be probing the interface between CGM associated with individual galaxies and the larger-scale network of intergalactic gas filaments, thus the lack of any correlation with azimuth angle is not wholly unexpected.

We do, however, detect an inclination effect on the density, and possibly $EW$, of Ly$\alpha$ absorbers. The combination of no azimuthal dependence and increased absorber density with inclination leads us to conclude that these galaxies have disk-like, oblate-spheroidal halos. A perfectly spheroidal halo would show no correlation for either, and an extremely flattened halo would show up as enhanced number density along the major axis. These results are consistent with a picture where the H\,{\sc i} covering fraction steadily decreases from $\sim$unity very near to galaxy disks out to at least 1 Mpc, where gas associated with galaxies merges with the general IGM. Our larger upcoming dataset will provide the statistics necessary to probe this in finer detail, as well as give clues regarding the exact shape of this fall-off and the level of clumpiness or filamentary structure in galaxies' H\,{\sc i} halos.

\begin{table}[b]\footnotesize
  \caption{\small{Average properties of the associated galaxy sample split into red and blue-shifted bins based on $\Delta v$}}
  \vspace{-20pt}
\begin{center}
\begin{tabular}{l l l}
 \hline \hline
 Statistic                				&  Blueshifted Absorbers   &     Redshifted Absorbers     \\ 
  \hline \hline
 Number 	          			 		&     	22				&	26			\\
 Mean $EW$    \scriptsize $\rm [m\AA]$    &	$329 \pm 52$ 		&	$245 \pm 34$  	\\
 Median EW     \scriptsize $\rm [m\AA]$    & 	$292 \pm 16$		& 	$177 \pm 10$	\\
 Mean $\rm R_{vir}$   \scriptsize [kpc]	&   	$215 \pm 10$		& 	$224 \pm 10$	\\
 Mean $\rho$   \scriptsize [kpc]          		&   	$218 \pm 17$ 		& 	$298 \pm 23$	\\
 Mean $\Delta v$  \scriptsize $\rm [km\, s^{-1}]$     &	$ 68 \pm 16$    &	$-108 \pm 20$	\\
 Mean Inc.  \scriptsize [deg]  			&  	$58 \pm 4$		&	$61 \pm 4$	\\
 Mean Az.  \scriptsize [deg]    			&	$43 \pm 5$ 		&	$49 \pm 5$ 	\\
 
\hline
\end{tabular}

\footnotesize \raggedright \textbf{Note.} All reported errors are standard errors in the mean.
\end{center}
\label{resultsTable}
\end{table}

\vspace{5pt}

\section{SUMMARY}

We have introduced a novel likelihood method for associating absorption systems with nearby galaxies, and explored its implementation with a small subsample of 33 COS sightlines. Associating CGM absorbers with individual galaxies remains a difficult and ambiguous affair, but with this new metric we can at least do so in a reproducible and numerical manner. 

In this pilot sample we have measured 48 $\rm Ly\alpha$ absorption lines in the spectra of 33 COS targets and matched each to a single, large ($D\geq 25$ kpc) galaxy. Table \ref{resultsTable} presents a breakdown of our results when separating absorber-galaxy pairs into red and blue-shifted samples. The following summarizes our findings:

\vspace{10pt}

\begin{enumerate}

\item We introduce a likelihood parameter, $\mathcal{L}$, based on Gaussian profiles centered around $\rho / R_{eff}$ and $\Delta v / v_{norm}$ to automate the matching of absorbers with associated galaxies. The response of $\mathcal{L}$ can be tailored by choosing different values for $R_{eff}$ and $v_{norm}$ (we used $R_{eff}$ = [$R_{vir}$, $D^{1.5}$] and $v_{norm}$ = 200 $\rm km\, s^{-1}$ in this work, and will explore other parameterizations in a future paper).

\item Equivalent width ($EW$) anti-correlates most strongly with $\rho$ when normalized by $R_{vir}$. It follows that $EW$ weakly correlates and anti-correlates with $R_{vir}$ and $\rho$, respectively.

\item The mean and maximal $EW$ of absorbers increase with decreasing $\Delta v$. The strongest absorbers are nearly all found within $\Delta v = \pm 100$ $\rm km\, s^{-1}$ of their associated galaxies.

\item $\rm Ly\alpha$ absorbers are most commonly associated with inclined galaxies. $73\%$ of blueshifted and $77\%$ of redshifted absorbers are associated with galaxies with $i \geq 50$ deg, whereas $56\%$ of all galaxies in the survey volume have similarly high inclinations. The distributions of associated versus all galaxy inclinations differ at a greater than $99\%$ confidence, or $\sim 3.6\sigma$ level, according to the Anderson-Darling distribution test.

\item We find no strong evidence for azimuth preference for absorption - Ly$\alpha$ absorbers appear to be distributed uniformly around galaxy major and minor axes.

\end{enumerate}

In a future paper we will apply this method to a sample of hundreds of COS sightlines in an effort to produce the most statistically robust CGM study to date.

\acknowledgements

We would like to thank the referee for his/her valuable comments and suggestions. This research has made use of the NASA/IPAC Extragalactic Database (NED) which is operated by the Jet Propulsion Laboratory, California Institute of Technology, under contract with the National Aeronautics and Space Administration. This work is based on observations with the NASA/ESA \textit{Hubble Space Telescope}, obtained at the Space Telescope Science Institute (STScI), which is operated by the Association of Universities for Research in Astronomy, Inc., under NASA contract NAS 5-26555. Spectra were retrieved from the Barbara A. Mikulski Archive for Space Telescopes (MAST) at STScI. Over the course of this study, D.M.F. and B.P.W. were supported by grant AST-1108913, awarded by the US National Science Foundation, and by NASA grants \textit{HST}-AR-12842.01-A, \textit{HST}-AR-13893.01-A, and \textit{HST}-GO-14240 (STScI).

\facility{HST (COS)}

\nocite{*}
\bibliography{french_bib2}
\bibliographystyle{aasjournal}

\end{document}